# Developing and Evaluating Deep Neural Network-based Denoising for Nanoparticle TEM Images with Ultra-low Signal-to-Noise

Joshua L. Vincent[1], Ramon Manzorro[1], Sreyas Mohan[2], Binh Tang[3], Dev Y. Sheth[4], Eero P. Simoncelli[2,5,6], David S. Matteson[3], Carlos Fernandez-Granda[2,6]*, and Peter A. Crozier[1]*

[1] *School for Engineering of Matter, Transport, and Energy, Arizona State University, Tempe, AZ. 85287*

[2] *Center for Data Science, New York University, New York City, NY 10011*

[3] *Department of Statistics and Data Science, Cornell, Ithaca, NY 14850*

[4] *Indian Institute of Technology Madras, Chennai, Tamil Nadu 600036, India*

[5] *Center for Neural Science, New York University, New York City, NY 10011*

[6] *Courant Institute of Mathematical Sciences, New York University, New York City, NY 10011*

*Corresponding author emails: cfgranda@cims.nyu.edu and crozier@asu.edu

**Abstract**

A deep convolutional neural network has been developed to denoise atomic-resolution TEM image datasets of nanoparticles acquired using direct electron counting detectors, for applications where the image signal is severely limited by shot noise. The network was applied to a model system of $CeO_2$-supported Pt nanoparticles. We leverage multislice image simulations to generate a large and flexible dataset for training the network. The proposed network outperforms state-of-the-art denoising methods on both simulated and experimental test data. Factors contributing to the performance are identified, including (a) the geometry of the images used during training and (b) the size of the network's receptive field. Through a gradient-based analysis, we investigate the mechanisms learned by the network to denoise experimental images. This shows that the network exploits global and local information in the noisy measurements, for example by adapting its filtering approach when it encounters atomic-level defects at the nanoparticle surface. Extensive analysis has been done to characterize the network's ability to correctly predict the exact atomic structure at the nanoparticle surface. Finally, we develop an approach based on the log-likelihood ratio test that provides a quantitative measure of the agreement between the noisy observation and the atomic-level structure in the network-denoised image.

## 1. Introduction

Even with a perfect electron detector, Poisson noise degrades the information content of an atomic resolution electron microscope image, lowering the sensitivity for atomic column detection and limiting the precision for determining the atomic column occupancy. The Poisson statistics of an image can be improved by counting for longer times or by increasing the electron beam current, although this is not always possible. In beam sensitive systems such as organic materials or liquids, extended electron irradiation induces undesirable changes in the structure and composition of the sample. Additionally, for investigation of dynamic processes with time-resolved *in situ* microscopy, the short exposure time per frame may result in very low signal-to-noise (SNR) values. An example of an important class of nanomaterials where dynamic structural changes may strongly influence functionality is heterogeneous catalysts. One approach to address this SNR challenge is to develop denoising techniques which effectively estimate and partially restore some of the information missing from the experimental image. The details and effectiveness of such approaches to atomic resolution electron microscopy images have not been well explored. Here, we develop and evaluate deep learning methods for denoising the images of catalytic nanoparticle surfaces recorded from aberration-corrected transmission electron microscope. While our primary motivation is catalysis, the approaches described here may be applicable to a wider range of atomic resolution imaging applications that are characterized by ultra-low SNR.

Heterogeneous catalysts are an important class of materials due to their immense impact on energy and the environment. Aberration-corrected *in situ* environmental transmission electron microscopy (ETEM) is a powerful tool capable of providing atomic-scale information from technical catalysts under reaction conditions (Crozier & Hansen, 2015; Tao & Crozier, 2016; Dai et al., 2017; He et al., 2020). Highly resolved, atomic-level information is vital to establish improved catalyst design principles, as it is now well known that catalytic functionality is governed by surface sites which typically consist of only a few atoms (Nørskov et al., 2014; Schlögl, 2015).

Catalytically relevant structures may only form under reaction conditions and can continuously reconfigure as a result of interactions with adsorbates and reaction intermediates (Vendelbo et al., 2014; Bergmann & Roldan Cuenya, 2019). The importance of such dynamic or so-called fluxional behavior was recognized many years ago in the surface science (Somorjai, 1991) and chemistry (Cotton, 1975) communities, and is becoming increasingly recognized theoretically (Zhai & Alexandrova, 2017; Zhang et al., 2020) and experimentally (Lawrence et al., 2021; Li et al., 2021) as critical for understanding the functionality of catalytic nanomaterials. Recent advancements in the realization of highly efficient direct electron detectors now enable atomically-resolved ETEM image time series to be acquired with a temporal resolution in the millisecond (ms) regime (Faruqi & McMullan, 2018; Ciston et al., 2019). As many catalysts exhibit chemical reaction turnover frequencies on the order of $10^0 - 10^2$ sec$^{-1}$, the emerging opportunity to visualize atomic-level behavior with high temporal resolution holds much promise for understanding the chemical transformation processes that take place on catalyst surfaces.

Although there is potentially much to be gained from applying these new detectors to catalytic nanomaterials characterization, acquiring *in situ* TEM image time series with very high temporal resolution produces datasets that can be severely degraded by noise (Lawrence et al., 2020). Cutting-edge sensors offer detective quantum efficiencies approaching the theoretical maximum of unity, largely by eliminating readout noise and employing electron counting to significantly improve the modulation transfer function (Ruskin et al., 2013; Faruqi & McMullan, 2018). Even so, especially at high speeds, where the average dose is often < 1 e$^-$ per pixel per frame, the information content of the image signal still remains limited by fundamental Poisson shot noise that is associated with the electron emission and scattering processes.

Following Poisson statistics, counted images with an average dose < 1 e$^-$/pixel have signal-to-noise ratios (SNR) on the order of unity, and consequently, ascertaining the underlying structure in the image becomes a major obstacle. By carefully selecting and summing frames in a time-series, precise structural information on metastable states can be obtained with improved SNR. Averaging consecutive frames may reveal fluxional behavior provided the lifetime of each metastable state is longer that the averaging time. However, information on short-lived intermediate states, *which may ultimately underpin the material's overall functionality*, may no longer be resolvable at larger temporal resolutions.

Thus, there is a pressing need for sophisticated noise reduction techniques that preserve the temporal resolution of the image series and facilitate the retrieval of features at the catalyst surface. Convolutional neural networks (CNNs) achieve state-of-the-art denoising performance on natural images (Liu & Liu, 2019; Tian et al., 2019; Zhang et al., 2017) and are an emerging tool in various fields of scientific imaging, for example, in fluorescence light microscopy (Zhang et al., 2019; Belthangady & Royer, 2019) and in medical diagnostics (Yang et al., 2017; Jifara et al., 2019). In electron microscopy, deep CNNs are rapidly being developed for denoising in a variety of applications, including structural biology (Buchholz et al., 2019; Bepler et al., 2020), semiconductor metrology (Chaudhary et al., 2019; Giannatou et al., 2019), and drift correction (Vasudevan & Jesse, 2019), among others (Ede & Beanland, 2019; Lin et al., 2021; Spurgeon et al., 2021), as highlighted in a recent review (Ede, 2020).

To our knowledge, deep neural networks have not yet been developed to denoise atomic-resolution TEM images of catalyst nanoparticles with an emphasis on atomic scale surface structure. As the potentially fluctuating atomic-scale structure at the catalyst surface is of principal scientific interest in this application, it is critical to establish methods for evaluating the agreement between the noisy observation and the structure that appears in the network-denoised image. As far as we are aware, such analysis is not found in the previous literature on CNNs for electron-micrograph denoising. Moreover, the mechanisms by which trained networks successfully denoise is often treated as a "black box". Revealing and studying these mechanisms is, however, a key step towards further improving this methodology and understanding its potential and limitations.

In this paper, we develop a supervised deep CNN to denoise atomic-resolution TEM images of nanoparticles acquired in applications where the image signal is severely limited by shot noise. The network was trained on a dataset of simulated images produced through multislice calculations and then applied to experimentally acquired images of a model system which consists of $CeO_2$-supported Pt nanoparticles. We perform an extensive analysis to characterize the network's ability to recover the exact atomic-scale structure at the Pt nanoparticle surface. We also establish an approach to assess the agreement between the noisy observation and the atomic structure in the network-denoised image, without access to ground-truth reference images. Finally, we investigate the mechanisms used by the network to denoise experimental images and present a visualization of these mechanisms in the form of equivalent linear filters, which reveal how the network adapts to the presence of non-periodic atomic-level defects at the nanoparticle surface.

## 2. Materials and Methods

### 2.1. Experimental Data Acquisition

Atomic-resolution image time-series of $CeO_2$-supported Pt nanoparticles were acquired to provide experimental data for testing and developing the denoising network. Acquiring image time series at high speed is one application that results in ultra-low SNR images and is thus an appropriate focus for the methodological development described here. The nanoparticles were synthesized through standard hydrothermal and metal deposition methods that have been described previously (Vincent & Crozier, 2019). Time-resolved series of images were acquired on an aberration-corrected FEI Titan ETEM operated at 300 kV. The third-order spherical aberration coefficient (C3) of the aberration corrector was tuned to a slightly negative value of approximately -13 μm, yielding a white-column contrast for the atomic columns in the resultant images. The measured 5$^{th}$ order spherical aberration coefficient (C5) was 5 mm. Lower-order aberrations, e.g., astigmatism and coma were continuously tuned to be as close to 0 nm as possible and thus considered to be negligible. TEM samples were prepared by dispersing the Pt/CeO$_2$ powder onto a windowed micro electro-mechanical system (MEMS)-based Si$_3$N$_4$ chip. After loading the sample into the ETEM, nitrogen gas was leaked into the cell until an ambient pressure of 5 x 10$^{-3}$ Torr N$_2$ was achieved; the temperature was maintained at 20 °C. It is briefly mentioned that this dataset is part of a larger series of images of the same catalyst imaged in N$_2$ and under a CO oxidation gas atmosphere, wherein the catalyst exhibits very rapid structural dynamics that present considerable modeling challenges (Vincent & Crozier, 2020). Hence, for this work, the image time-series of the catalyst in a N$_2$ atmosphere was chosen to provide a practicable starting point for developing the network as well as for assessing its performance. Time-resolved image series were acquired using a Gatan K2 IS direct electron detector. Images were taken at a speed of 40 frames per second (fps), yielding a time resolution of 25 milliseconds (ms) per frame. An incident electron beam dose rate of 5,000 e$^-$/Å$^2$/s was used; for the pixel size employed during the experiment (i.e., 0.061 Å /pixel), these conditions resulted in an average dose of 0.45 e$^-$/pixel/frame. The frames of the time-series were aligned without interpolation after acquisition. The electron beam was blanked when images were not being acquired.

*2.2. Atomic Model Generation and Multislice TEM Image Simulation*

A crucial step to achieve effective denoising performance with the supervised deep convolutional neural network is to carefully design the training dataset. Here, a wide range of structural configurations and imaging conditions were pursued (a) to encompass potential variations that could occur experimentally and (b) to explore the effect of training and testing the network on various subsets of images generated under different conditions. In all, we have produced 17,955 image simulations of Pt/$CeO_2$ models by systematically varying multiple imaging parameters and specimen structural configurations, e.g., defocus, tilt, thickness, the presence of surface defects, Pt nanoparticle size, etc. The 3D atomic structural models utilized in this work consist of Pt nanoparticles that oriented in a [110] zone axis and that are supported on a $CeO_2$ (111) surface which is itself oriented in the [110] zone axis. This crystallographic configuration corresponds to that which is often observed experimentally and is thus the focus of the current work. The models have been constructed with the freely available Rhodius software (Bernal et al., 1998). The faceting and shape of the supported Pt nanoparticle was informed by surface energies reported by McCrum et al (McCrum et al., 2017). A Wulff construction based on these values was built in the MPInterfaces Python package (Mathew et al., 2016) and iteratively adjusted in size until a qualitative match in dimension was achieved with the experimentally-observed shape.

A total of 855 atomic-scale structural models of Pt/$CeO_2$ systems were created. Each model represents Pt nanoparticles of various size, shape, and atomic structure (e.g., small, medium, or large size, with either faceted or defected surfaces, or some combination of both), supported on $CeO_2$, which itself may present either a faceted surface or one characterized by surface defects. Extended details on the modeled structures are given in **Supplemental Appendix A**. Each model consists of a supercell having *x* and *y* dimensions of 5 nm x 5 nm. The support thickness was systematically varied between 3 nm and 6 nm in 1 nm increments, so the supercell's $z$ dimension varies depending on the thickness of the particular model. The orientation of the structure with respect to the incident electron beam was also systematically varied from 0° to 4° about the $x$ and $y$ axes independently in increments of 1°. Thus, variations from 0° in $x$ and 0° in $y$, to 4° in $x$ and 0° in $y$, or 0° in $x$ and 4° in $y$ were considered.

Simulated HRTEM images were generated using the multi-slice image simulation method, as implemented in the Dr. Probe software package (Barthel, 2018). Given the low pressure of gas present during the experimental image acquisition (i.e., < 1 Pa), the presence of $N_2$ was ignored during the image calculation, which is supported by experimental measurements done by Hansen and colleagues (Hansen & Wagner, 2012). All of the simulations were performed using an accelerating voltage of 300 kV, a beam convergence angle of 0.2 mrad and a focal spread of 4 nm. A slice thickness of 0.167 Å was used. Following the experimental conditions, the third-order spherical aberration coefficient (Cs) was set to be -13 μm. The fifth-order spherical aberration coefficient (C5) was set to be 5 mm. All other aberrations (e.g., 2-fold and 3-fold astigmatism, coma, star aberration, etc.) were approximated to be 0 nm. To make the process of computing nearly 18,000 image simulations tractable, the calculations were performed in a parallel fashion on a supercomputing cluster (Agave cluster at ASU).

To explore the effect of defocus on the training and testing of the network, the defocus value (C1) was varied from 0 nm to 20 nm in increments of 1 nm. Image calculations were computed using a non-linear model including partial temporal coherence by explicit averaging and partial spatial coherence, which is treated by a quasi-coherent approach with a dampening envelope applied to the wave function. An isotropic vibration envelope of 50 pm was applied during the image calculation. Images were simulated with a size of 1024 x 1024 pixels and then later binned with cubic interpolation to desired sizes to match the pixel size of the experimentally acquired image series. Finally, to equate the intensity range of the simulated images with those acquired experimentally, the intensities of the simulated images were scaled by a factor which equalized the vacuum intensity in a single simulation to the average intensity measured over a large area of the vacuum in a single 25 ms experimental frame (i.e., 0.45 counts per pixel in the vacuum region).

To exemplify the variation incorporated into the overall training dataset, **Figure 1a** depicts a representative subset of four Pt/CeO$_2$ atomic structural models, along with (**Figure 1b**) three randomly selected multislice TEM image simulations generated from each model. The structural models are shown in two perspectives: a tilted view to emphasize 3D structure (first column), and a projected view along the electron beam direction (second column). Note the variation in Pt particle size, shape, and surface defect structure, as well as the changes to the CeO$_2$ support surface character, with the bottom model displaying a Pt particle with a single atom surface site along with a CeO$_2$ support having multiple step-edge defects. Accounting for the remaining particle and support structures, in addition to the variations in crystal orientation and CeO$_2$ support thickness, a total of 855 such models were constructed. These structures were each used to calculate multislice simulations with 21 defocus values incremented from 0 to 20 nm in 1 nm intervals, which results in the calculation of $855 \times 21 = 17,955$ total images. Simulations randomly selected from each model and shown in **Figure 1b** demonstrate the large variety of signal contrast and specimen structure available for training and testing the neural network.

*2.3. Convolutional Neural Network Training and Testing*

Before being applied to the real data, the networks were trained and evaluated on various subsets of simulated images. As will be discussed below, typically around 5,500 simulated images were used to train the network, with 550 other images randomly selected for validation and testing. Noisy data for training and evaluating the network were generated from clean simulated images by artificially corrupting the clean simulations with Poisson shot noise. That is, a noisy simulated image was produced pixel-wise by randomly sampling a Poisson distribution with a mean value equal to the intensity in the corresponding pixel of the clean ground truth image. We have verified that the noise in the experimental counted TEM image time-series follows a Poisson distribution (**see Supplemental Appendix B** and **Figure S15**), which is expected given the physical origin of the shot noise in the electron counted image acquisition process.

The network training process involves (1) denoising a noisy image, (2) comparing the denoised output to the clean ground truth through a quantitative loss function, and (3) adjusting the parameters of the network iteratively to achieve better performance. The parameters are adjusted via back-propagation using the stochastic gradient descent algorithm (Goodfellow et al., 2016). Periodically the network is evaluated on a validation set of images not included in the training set. We chose to quantify the difference between the output and the ground truth by computing the L2 norm or mean squared error (MSE) of the two images, as is standard in the denoising literature. The magnitude of this value is conveniently represented by a related quantity known as the peak signal-to-noise ratio, or PSNR, which can be calculated from the MSE by the following equation:

(1) $\quad PSNR = 10 \times \log_{10}(\frac{MAX_I^2}{MSE})$

Here, $MAX_I$ is the maximum possible intensity value of the clean ground truth image. The PSNR is essentially a decibel-scale quantity that is inversely proportional to the MSE: a very noisy image will have a low PSNR. The PSNR for the noisy images in this work is around 3 dB.

It is desirable to investigate the performance of the network when applied to images that differ from the those that were present in the training data. To evaluate this so-called generalization ability, we divided the simulated dataset into various subsets, systematically trained the network on one of the subsets, and then evaluated its performance on the rest. In these cases, the number of images in each training subset was set equal to establish a fair assessment. The subsets were classified based on (1) the character of the atomic column contrast, (2) the structure/size of the supported Pt nanoparticle, and (3) the non-periodic defects present in the Pt surface. The atomic column contrast was classified into three divisions: black, intermediate, or white, largely based on the Pt and Ce atomic column intensities (see, e.g., **Supplemental Figure S1**). The nanoparticle structures were classified into four categories, "PtNp1" through "PtNp4", each with different size and shape, in accordance with the models displayed in **Supplemental Figure S11**. Finally, the defects were divided into five categories: "D0", "D1", "D2", "Dh", and "Ds", in accordance with the models presented in **Supplemental Figure S12**.

All networks (e.g., the proposed architecture as well as those used in the baseline evaluation methods described below) were trained on 400 x 400 pixel sized patches extracted from the training images and augmented with horizontal flipping, vertical flipping, random rotations between -45° and +45°, as well as random resizing by a factor of 0.955 – 1.055. The models were trained using the Adam optimizer (Kingma & Ba, 2015), with a default starting learning rate of 1 $\times$ 10$^{-3}$, which was reduced by a factor of two each time the validation PSNR plateaued. Training was terminated via early stopping, based on validation PSNR (Goodfellow et al., 2016).

The proposed network architecture is a modified version of U-Net (Ronneberger et al., 2015) with six scales to achieve a large field of view (roughly 900 x 900 pixels). The network consists of 6 down-blocks and 6 up-blocks. A down-block consists of a max-pooling layer, which reduces the spatial dimension by half, followed by a convolutional-block (conv-block). Similarly, an up-block consists of bilinear up-sampling, which enlarges the size of the feature map by a factor of two, followed by a conv-block. Each conv-block itself consists of conv-BN-ReLU-conv-BN-ReLU, where conv represents a convolutional layer, BN represents a batch normalization process (Ioffe & Szegedy, 2015), and ReLU represents a non-linear activation by a rectified linear unit. In our final model, we use 128 base channels in each layer of conv-block.

*2.4. Baseline Methods for Denoising Performance Evaluation*

A number of other methods, including other trained denoising neural networks that are typically applied to natural images, were also applied both to the simulated and the real data in order to establish a baseline for evaluating the performance of the proposed network. A brief overview of the methods will be given here. The performances of the methods were compared quantitatively in terms of PSNR and structural similarity (SSIM) (Wang et al., 2004).

(a) **Adaptive Wiener Filter (WF):** An adaptive low-pass Wiener filter was applied to perform smoothing. The mean and variance of each pixel were estimated from a local circular neighborhood with a radius equal to 13 pixels.

(b) **Low-pass Filter (LPF):** A linear low-pass filter with cut-off spatial frequency of 1.35 Å$^{-1}$ was applied to preserve information within the ETEM instrumental resolution while discarding high-frequency noise.

(c) **Variance Stabilizing Transformation (VST) + Non-local Means (NLM) or Block-matching and 3D Filtering (BM3D):** NLM and BM3D are commonly used denoising routines for natural images with additive Gaussian noise (Buades et al., 2005; Makitalo & Foi, 2013). Here, a nonlinear VST (the Anscombe transformation) was used to convert the Poisson denoising problem into a Gaussian denoising problem (Zhang et al., 2019). After applying the Anscombe transformation, we apply BM3D or NLM to the transformed image, and finally use the inverse Anscombe transformation to recover the denoised image.

(d) **Poisson Unbiased Risk Estimator + Linear Expansion of Thresholds (PURE-LET):** PURE-LET is a transform-domain thresholding algorithm adapted to mixed Poisson-Gaussian noise (Luisier et al., 2011). The method requires the input image to have dimensions of the form $(2^n, 2^n)$. To apply this method here, $128 \times 128$ pixel-sized overlapping patches were extracted from the image of interest, denoised individually, and finally stitched back together by averaging the overlapping pixels.

(e) **Blind-spot Denoising:** We trained a blind-spot network based on U-net, as developed by Laine et al. (Laine et al., 2019). Here, training was done using $600 \times 600$ pixel-sized patches from the images of interest. The Adam optimizer was used with a starting learning rate of $1 \times 10^{-4}$, which was reduced by a factor of two every 2,000 epochs. Overall, the training proceeded for a total of 5,000 epochs.

(f) **Denoising Convolutional Neural Network (DnCNN):** Following the protocol outlined in the Section 2.3, we trained the DnCNN model as described previously by Zhang and coworkers (Zhang et al., 2017).

(g) **Small U-Net from Dynamically Unfolding Recurrent Restorer (DURR):** Following the protocol outlined in the Section 2.3, we trained a U-Net architecture implemented in the DURR denoiser proposed by Zhang and coworkers (Zhang et al., 2018).

Aside from these methods, standard filtering techniques including Gaussian blurring, median filtering, and Fourier transform (FT) spot-mask filtering were applied using routines built-in to the ImageJ analysis software (Schneider et al., 2012). Where relevant, additional details will be given to aid in understanding.

## 3. Results and Discussion

*3.1. Need for Improved Denoising Methods and Overview of CNN -based Deep Learning Denoiser*

A single 25 ms exposure counted frame of a CeO$_2$-supported Pt nanoparticle from an experimentally-acquired time-resolved *in situ* TEM image series is presented in **Figure 2(a1 and a2)**. The Pt particle is in a [110] zone axis on a [111] CeO$_2$ surface that is itself in a [110] zone axis orientation. These orientation relationships and particle/support zone axes were commonly encountered during the experiment. Even though a relatively high dose rate of $5 \times 10^3$ e$^-$/Å$^2$/s was used to acquire the image series, for time-resolved frame rates on the order of ms, many of the pixel values are zero. In the present case, the average electron dose counted in the vacuum region of the image is 0.45 e$^-$/pixel/frame. Following Poisson statistics, wherein the standard deviation of the signal is equal to the mean value, and assuming the intensity in the vacuum region is uniform, the signal-to-noise ratio (SNR) of the incident beam is only $SNR = \frac{0.45}{\sqrt{0.45}} = \sqrt{0.45} = 0.67 < 1$. Hence, the image is severely degraded by shot noise. The impact of the shot noise limitation is emphasized by magnifying the region marked by the dashed white box, which is presented in **Figure 2(a2)**. Here, the quality of the signal is appreciably low, and the Pt atomic columns at the nanoparticle surface are hardly discernible.

One common approach to improving the SNR of time-resolved image series involves aligning and then summing together non-overlapping groups of sequential frames, yielding a so-called time-averaged or summed image. **Figure 2(b1)** presents a 1.000 sec time-averaged image produced from adding together 40 sequential 0.025 sec frames. The pronounced improvement in SNR, which has increased by a factor of $\sqrt{40} = 6.32$ to $SNR = 4.24$, is readily evident, as seen by the well-defined and bright atomic columns that appear in **Figure 2(b2)**.

Increasing the SNR without time-averaging can be accomplished by applying linear or non-linear filters that act on variously sized and/or distributed domains in real or frequency space to remove sharp features arising from high noise content. The result of applying a non-linear median filter with a 3 × 3 pixel-sized kernel to the noisy single frame is presented in **Figure 2c**. The application of a linear Gaussian blur with a kernel that has a standard deviation equal to 1 pixel yields the filtered image presented in **Figure 2d**. Applying kernels of these size and character produced the best improvement in image quality for each filter. Although the filtered images appear smoother and offer an enhanced visualization of the atomic columns in comparison to the raw image, the action of the filters also introduces artifacts to the signal, which can complicate a precise analysis of the atomic column position and/or intensity.

Working in reciprocal space through the application of a Fourier transform (FT) allows one to consider spatial frequency filters that exclude components attributable to noise, with a subsequent reconstruction of the image using the desired domains from the filtered FT. **Figure 2(e1)** presents a Fourier reconstruction of the individual frame after applying a linear low-pass filter that excludes components with spatial frequencies beyond the instrument's 1.35 Å$^{-1}$ information limit. After eliminating the high frequency information corresponding to noise, the contrast in the image exhibits an unusual texture that hinders feature identification, as seen in **Figure 2(e2)**. **Figure 2f** displays another Fourier reconstruction produced here by spot-masking the regions corresponding to Bragg beams in the FT, as presented in the figure inset. Although this reconstructed image offers an improved SNR compared to the raw frame and even to the other filtering techniques, the procedure introduces severe ringing lattice-fringe artefacts into the vacuum region and at the nanoparticle surface, making it unacceptable for use in the study of defects or aperiodic structures.

There is a pressing need for improved denoising techniques that both preserve the high time resolution of the original data and also facilitate the retrieval of non-periodic structural features, e.g., nanoparticle surfaces and atomic-level defects. Toward this end, we develop a deep CNN that is trained on a big dataset of simulated TEM images before being applied to real data.

A schematic overview of the deep CNN training, application, and evaluation process is provided in **Figure 3**. During training (top), a large dataset of noisy simulated images is given to the network. Noisy images were generated from clean simulated images by corrupting them with Poisson shot noise. For each noisy image, the network produces a prediction of the underlying signal, effectively denoising the image. The denoised prediction is compared to the original clean simulation by computing the L2 norm, or mean squared error (MSE), between the two images. Better denoising performance is achieved by iteratively adjusting the parameters within the network (via stochastic gradient descent or a related optimization algorithm), in order to minimize the MSE difference between the denoised output and the original simulation. After successfully training the network, it may be applied to real data (bottom). The denoised experimental 25 ms frame produced by the network presents a significant improvement in SNR without temporal averaging and without making sacrifices to the study of non-periodic structural features. However, given the high level of noise present in the raw data, caution must nonetheless be exercised when performing analysis on the network denoised output. As will be shown, we have established an approach for quantifying the degree of agreement between the network estimated output and the noisy raw input, which takes the form of a statistical likelihood map.

*3.2. Performance of Trained Network on Validation Dataset of Simulated Images*

Before applying the trained network to real data, it is important to assess and validate the network's performance on noisy simulated data that it has not seen before. **Figure 4** presents a representative comparison of the surveyed methods against our proposed network on an image randomly selected from the validation dataset. A similar comparison for another randomly selected image in the validation dataset is given in **Supplemental Figure S2**. The aggregate performance, in terms of PSNR and structural similarity (SSIM, (Wang et al., 2004)), for each denoising approach over all images in the validation dataset is summarized in **Table 1**. Descriptions of each method are given in detail in **Section 2.4**. The noisy simulated image shown in **Figure 4a**, along with the zoom-in image taken from the region indicated by the red box along the Pt nanoparticle surface, illustrating the severity of the signal degradation that has occurred due to shot noise. The same noisy image was processed using the denoising methods described in **Section 2.4**. The results are presented from **Figure 4b** to **Figure 4i** in order of increasing performance in terms of PSNR. The original ground truth simulated image, which serves as a ground truth reference, is presented in **Figure 4j**.

In general, the proposed deep CNN denoising architecture outperforms the baseline methods by a large margin, achieving a PSNR of $42.87 \pm 1.45$ dB and a SSIM of $0.99 \pm 0.01$. The starting PSNR of the noisy simulation is about 3 dB. As seen in **Figure 4i**, the proposed network produces an estimated image that closely resembles the ground truth simulation. In addition to recovering the overall shape of the Pt nanoparticle, the aperiodic structures of the Pt surface and the Pt/$CeO_2$ interface, as well as the subtle contrast variations that are present in the $CeO_2$, have all been accurately denoised by the proposed architecture. The next best performance is attained by the other two simulation-based denoising (SBD) neural networks (e.g., **Figure 4g/h**), which reach

PSNR and SSIM values around 30.6 dB and 0.93, respectively. However, in the images denoised through these inferior networks, the contrast features around aperiodic sites or abruptly terminating surfaces are typically missing or distorted. Moreover, significant artifacts often appear in these images, including phantom atomic column-like contrast in the vacuum, or unrealistic structures characterized by missing columns in unphysical sites, e.g., the material bulk.

A number of decisive factors contribute to the performance of the network. First is the size of the network's receptive field. The receptive field is the region of the noisy image that the network can see while estimating the intensity of a particular denoised output pixel. The baseline networks included in the performance comparison, which are the present state-of-the-art in denoising natural images, employ receptive fields either $41 \times 41$ pixels (in the case of DnCNN, **Figure 4g**) or $45 \times 45$ pixels (in the case of the small UNet, **Figure 4h**). Given the fact that the real space pixel size of the data is 6.1 pm, these receptive fields amount to regions around 0.26 nm $\times$ 0.26 nm in size. As shown in **Supplemental Figure S3**, with a limited receptive field of such size, it is challenging to see the structure of the atomic columns in the ground truth simulation. Once shot noise has been added to reduce the PSNR to 3 dB, differentiating regions containing structure from those which contain only vacuum becomes virtually impossible by eye. Increasing the receptive field is critical to achieving better denoising performance. **Supplemental Figure S4** shows that expanding the receptive field by a factor of 25 to a region around $200 \times 200$ pixels (i.e., 1.22 nm $\times$ 1.22 nm) allows the network to sense the local structure around the pixel to be denoised. With a receptive field of this size, different structures (e.g., vacuum, Pt surface, $CeO_2$ bulk, surface corner site) remain discernible even after adding noise. This suggests that increasing the receptive field contributes to the network's ability to detect subtle contrast variations as well as aperiodic defects.

The receptive field of the proposed network is roughly 900 x 900 pixels (i.e., 5.49 nm $\times$ 5.49 nm, **Figure 4i**).

The network's performance is also influenced by the nature of the images contained in the training dataset. Here we have discovered that the geometry of the image (i.e., the scaling and orientation) as well as the character of the atomic column contrast (i.e., the focusing condition) appear to have the largest impact on performance. In **Supplemental Figure S5** we demonstrate that the denoising performance measured in terms of PSNR degrades significantly when the network is evaluated on simulated images that have been scaled or rotated in a manner that was missing from the images in the training dataset. Note that the performance remains roughly constant across various values of pixel size and orientation *when these pixel sizes and orientations are present in the training dataset.* These results indicate that augmenting the training data with random resizing/rotations can ensure that robust performance is obtained when the network is applied to real data, which may differ slightly in exact scaling or orientation from the images in the training dataset. Practically, the results also imply that networks must be carefully trained to denoise images taken at the particular image magnification of interest.

We have also investigated the generalizability of the network to unseen supported nanoparticle structures, non-periodic surface defects, and atomic column contrast conditions (i.e., defocus). As shown in **Supplemental Figure S6**, the network generalizes well to new (a) nanoparticle structures of various shape/size and (b) atomic-level Pt surface defects, with a good and consistent PSNR denoising performance above 34 dB for all of the categories explored here. The network is also generally robust to ± 5 nm variations in defocus. The largest degradation in performance (PSNR = 28 dB) is observed when the network is trained on images with black-column contrast and tested on images with white-column contrast. A general conclusion would be to train the network using images simulated at a defocus close to the data that is to be denoised.

*3.3. Evaluating the Network's Ability to Accurately Predict Nanoparticle Surface Structure*

Understanding the atomic-scale structure of the catalyst surface is of principal scientific interest. Here, we perform a detailed evaluation of the network's ability to produce denoised images that accurately recover the atomic-level structure of the supported Pt nanoparticle surface. The analysis was conducted over a set of 308 new simulated images that were specifically generated for the surface structure evaluation. A series of 44 Pt/$CeO_2$ structural models were created with many different types of atomic-level surface defects, including, e.g., the removal of an atom from a column, the removal of two atoms, the removal of all but one atom, the addition of an adatom at a new site, etc., to emulate dynamic atomic-level reconfigurations that could potentially be observed experimentally. Nine of the models are shown in **Supplemental Figure S7** to provide an overview of the type of surface structures that were considered. Images were simulated under defocus values ranging from 6 nm to 10 nm, all with a tilt of 3° in *x* and -1° in *y* and a support thickness of 40 Å. Note that these images were never seen by the network during the training process and demonstrate an evaluation of its performance on unseen images.

A ground truth simulated image from the surface evaluation dataset is shown in **Figure 5(a1)**. A so-called blob detection algorithm based on the Laplacian of Gaussian approach was implemented to locate and identify the Pt atomic columns in the image (Kong et al., 2013). The atomic columns at the nanoparticle surface were distinguished from those in the bulk, which was done by computing a Graham scan on the identified structure (Graham, 1972). **Figure 5(a2)** shows a binary image depicting the Pt atomic columns identified in the ground truth simulated image. The set of atomic columns located at the surface have been highlighted with a green line.

Evaluating the network's ability to recover surface structure can be accomplished by examining how this set changes after denoising. **Figure 5(b1)** displays a denoised image produced by the network from a unique noise realization of the ground truth simulation. While the network denoises with outstanding performance and recovers the overall shape of the specimen, note the appearance of the three spurious Pt surface atomic columns that do not appear in the original ground truth simulation. The Pt atomic columns identified in this denoised image are pictured in **Figure 5(b2)**, where those located at the surface are highlighted now by a red line. The spurious Pt surface atomic columns have been marked with white arrows. Based on inspection of the noisy data, we believe that the particular distribution of intensity present in the noise realization can lead the network to produce denoised estimates with spurious surface atoms, perhaps due to the random clustering of intensity in a manner that appears to resemble an atom (see, e.g., **Supplemental Figure S8**). **Figure 5(c1)** displays a denoised image produced by the same network from a *second* unique noise realization. Note that in this case the Pt surface structure has been *recovered exactly*. The Pt atomic columns identified in this denoised image are pictured in **Figure 5(c2)** and are equivalent to those identified in the original simulation.

To quantify the network's performance in recovering the Pt atomic structure, we compute four metrics that are commonly employed in the field of machine learning: precision, recall, F1 score, and Jaccard index. These metrics are defined by the following equations:

(2) $\quad Precision = \frac{|A \cap B|}{|B|}$

(3) $\quad Recall = \frac{|A \cap B|}{|A|}$

(4) $\quad F1\ Score = 2 \times \frac{Precision \times Recall}{Precision + Recall}$

(5) $\quad Jaccard\ Index = \frac{|A \cap B|}{|A \cup B|}$

These metrics were calculated for both the surface and the bulk structure; when the metrics were calculated for the surface structure, $|A|$ represents the set of Pt atomic columns identified at the surface in the ground truth simulation, and $|B|$ represents the columns identified at the surface in the denoised image. Similarly, when the metrics were calculated for the bulk structure (i.e., everything other than the surface), $|A|$ and $|B|$ represent the bulk atomic columns in the ground truth and denoised images, respectively. To attain an accurate representation of the network's performance, 25 noise realizations of each ground truth simulation were sampled and then denoised, resulting in an evaluation over 7,700 total images.

**Figure 5d** displays box plot distributions of the four metrics computed over all 7,700 images for both the surface (blue boxes) and the bulk (orange boxes). Box plots, or box-and-whisker plots, are useful for graphically visualizing distributions of data on the basis of the quartiles that exist within the distribution. The quartiles are a set of three numerical values that divide the number of data points in the distribution into four roughly equally-sized parts; e.g., the $2^{nd}$ quartile is the median or mid-point of the dataset when the values are ordered from smallest to largest, the $1^{st}$ quartile lies halfway between the smallest value and the median, and the $3^{rd}$ quartile lies halfway between the median and the largest value. In the box-and-whisker plot, the box is drawn from the $1^{st}$ quartile ($Q_1$) to the $3^{rd}$ quartile ($Q_3$) with the median value represented by a line within this box. Whiskers, which are lines extending beyond the edges of the box, can be useful for describing the behavior of the data that falls in the upper or lower quartile of the distribution. Here we choose to follow a standard practice for drawing the whiskers: a distance equal to 1.5x the interquartile range (defined by $Q_3 - Q_1$) is drawn from each edge of the box; on the top of the box, for example, the largest value above $Q_3$ that lies within this distance is defined as the edge of the top whisker; similarly, the smallest value below $Q_1$ that lies within this distance is defined as the edge of the bottom whisker. Values beyond the edge of the whiskers are considered outliers; here, they are drawn as small solid diamonds. As seen in **Figure 5d**, the box plots for the bulk are all narrow and have median values of 1.0, which is expected given that the network was not seen to produce images characterized by unphysical bulk structures, such as, e.g., missing interior atomic columns.

The distributions for the surface structure are slightly more varied and reveal detailed information about the performance of the network. First, consider the distribution for the precision (left-most box plot in **Figure 5d**). The precision, or the positive predictive value, measures the fraction of real surface columns over all of the surface columns identified in the denoised image. Effectively, a lower precision value indicates that there are more false positives (i.e., spurious surface columns) in the denoised output. As a reference, consider a ground truth simulation in which there are originally 15 atomic columns present at the surface (e.g., **Figure 5(a1)**). The addition of one spurious surface column would result in a precision value of 0.93, while the addition of three columns would yield a precision value of 0.80. As seen in **Figure 5d**, the median precision value is 1.0 and the first quartile lies nearby at 0.93. Thus, the precision distribution shows the network frequently produces denoised images that do not contain spurious atomic columns; occasionally it will include one, and rarely it will add two or more.

In addition to including spurious atomic columns, the network may fail to recover the full structure, resulting in a real column that is absent from the denoised image. The prevalence of this event can be captured by the recall, which measures the fraction of real columns over all of the columns identified at the surface in the *clean ground truth image*. Effectively, a lower recall value indicates that there are more false negatives in the network denoised output, which means that columns which were originally present in the ground truth image are no longer present in the network denoised output. As presented in **Figure 5d**, the median recall value is also 1.0, with a

Page 24

distribution that is similar to – but narrower than – the precision. These values again indicate an impressive performance by the network. Interestingly, the slightly smaller distribution suggests that the network may tend to include spurious atomic columns more often than it fails to sense real atomic columns.

Taking the harmonic mean of the precision and recall yields the F1 score, which accounts both for false positives as well as false negatives. Here, the median value of the F1 score distribution is around 0.96, and the first quartile lies around 0.93. Given that the median precision and recall are both 1.0, it is not surprising that the F1 score distribution is also narrow and clustered around high values (i.e., greater than 0.90). Note that the harmonic mean of 1.0 (the median precision/recall) and 0.93 (the first quartile of both distributions) equals 0.96, which is the median F1 score. Thus, the F1 score reveals that while the network may occasionally include a spurious column or fail to include a real one, combinations of these errors occur less frequently.

Finally, we have computed the Jaccard index to gauge the exact degree of similarity between the surface structure in the clean and denoised images. As defined above, the Jaccard index equals the fraction of true positives (i.e., real columns) over the union of surface columns identified in both the clean and the denoised images. The ideal value of 1.0 occurs only when the exact atomic structure is recovered. In general, for the images in the surface evaluation dataset, the addition of a spurious atomic column would give a Jaccard index of 0.87, while the omission of a real column would give a value of 0.93. The distribution plotted in **Figure 5d** shows that the median Jaccard index value is 0.93 and that the first quartile lies at 0.87. Observe that the third quartile lies at 1.0, signaling that the network will achieve a perfect performance in recovering the precise atomic structure at the surface at least 25% of the time, *despite the extreme degree of signal degradation that has occurred due to shot noise*. The location of the first quartile at 0.87 indicates that at least 66% of the errors involve the addition or omission of only one atomic column. The remaining errors, which represent at most 25% of the total data, involve the addition and/or omission of more than one atomic column. Further studies implementing this approach could be done in the future to assess the effect that varying the noise level has on the network's ability to predict the atomic level surface structure exactly.

*3.4. Quantifying the Agreement between the Noisy Observation and the Network-Denoised Output*

When applying the trained network to real data, the atomic structure in the network denoised output cannot be compared to a clean ground truth image, since none is available. Establishing a tool to assess the likelihood of an atomic column's appearance in the network-denoised image would thus be of great utility. Here, we develop a statistical analysis based on the log-likelihood ratio test that makes it possible to hypothetically evaluate whether an atomic column in the denoised image is (1) likely to represent a true atomic column in the structure or (2) likely to be an artifact introduced by the denoising neural network. Additionally, a graphical visualization of the log-likelihood ratio is created in the form of a likelihood map. The log-likelihood ratio method requires only the network denoised image and the noisy input and is therefore extensible to real experimental data, where no clean ground truth references exist.

First, we validate the analysis on a large dataset of simulated images, for which the true atomic structures are exactly known. **Figure 6** depicts a representative (a) noisy and (b) denoised image from the simulated dataset discussed in the prior section. To compute the log-likelihood ratio and generate the likelihood map, the following procedure is implemented: first, an atomic column in the denoised image is located, e.g., through blob detection, as was done in the previous section (here, we focus on the Pt columns, although the method is generalizable to any area of interest so long as it can be identified in the denoised image). As a simplifying assumption, we model the intensity of the atomic column as a constant value, which is obtained by averaging over all the denoised pixels in the region $R$ identified by the blob detection algorithm. In **Supplemental Figure S9** we show that for these imaging conditions this is a good assumption, provided that the region $R$ is restricted to a limited area (e.g., radius < 0.7 Å) within the atomic column where the intensity is largely invariant.

Second, we compute the statistical likelihood, $L$, of observing the noisy data in $R$ of the input, assuming the true signal in this region is the constant value calculated from the denoised output. We know that the observed signal is governed only by shot noise, which can be modeled with a Poisson distribution. And furthermore, we assume that every pixel is mutually independent, so that the overall likelihood in $R$ is simply the product of the individual probabilities for each pixel $i$ in $R$. Mathematically, the likelihood calculation is then defined by the following equation:

(6) $\quad L(R) = \prod_{i \in R} p_\lambda(x_i)$

Where $x_i$ is the intensity of the $i$th noisy pixel in $R$, and $p_\lambda$ is a Poisson probability mass function characterized by a mean of $\lambda$, which is equal to the constant value calculated from the denoised output. Here, a higher likelihood value would indicate a better level of agreement between the denoised output and the noisy data. To assess instead whether the column is an artifact of the denoising network, we also compute the likelihood of observing the noisy data in $R$ with the true signal now represented by the constant value of the vacuum (i.e., $\lambda = 0.45$).

Comparing the relative magnitude of these two likelihood values allows one to consider the whether the atomic column is likely to be real or spurious. How consistent either hypothesis is with the noisy observation can be tested by taking the natural log of the likelihood ratio (also known as a log-likelihood ratio test). Considering, e.g., the noisy and denoised images of **Figures**

**6a and 6b**, the results of this test are conveniently visualized for every atomic column detected in the denoised image through the likelihood map that is presented in **Figure 6c**. Positive (red) log-likelihood ratio values indicate the detected column is more consistent with the noisy data than is the presence of vacuum. Conversely, sites with negative (blue) values are less consistent with the data and may therefore be spurious additions. A spurious atomic column appears in this denoised image at the corner site marked by the black arrow on the left side of particle. Observe that the likelihood map displays a relatively large negative value of -0.012 at this site, signaling that the detected atomic column is inconsistent with the noisy data and likely to be a spurious column.

It should be discussed that the likelihood map shows a handful of sites that correspond to real atomic columns, but which nonetheless have negative log-likelihood ratio values, including, e.g., in the bulk of the nanoparticle. First, we point out that the likelihood map does not provide an absolute validation of the structure present in the denoised image but rather offers a visualization of the statistical agreement between this structure and the noisy input. In this case, the observed image has been so degraded by shot noise (vacuum SNR = 0.67) that, inevitably, a few real atomic columns will be observed to have average noisy intensities that are more consistent with the vacuum level. The sensitivity of the log-likelihood ratio in response to the overall SNR has not been investigated and could be the subject of future work. As a second point, the appearance of real atomic columns with negative log-likelihood ratio is in some way a testament to the network's ability to infer the presence of structure in spite of a SNR so low that the data appears more consistent with vacuum. This point is explored further in **Section 3.5**. It is also worth pointing out that in a time series of images, one would be able to look at the variation in the likelihood map for different frames to facilitate a more correct interpretation.

Regardless of these nuances, some useful heuristics may still be established that allow one to use the likelihood map to quickly assess the atomic structure that appears in the denoised image. **Figure 6d** presents letter value or so-called boxen plots of ~65,000 log-likelihood ratio values calculated over 1,540 denoised images (5 unique noise realizations of 308 ground truth images), providing insight into how the distribution of values derived from spurious atomic columns (top) compares with that derived from real atomic columns (bottom). A dashed vertical line is provided at 0.0 for reference. The spurious column distribution shows a slightly negative median and is clustered around 0.0 while being skewed toward negative values. The positive tail diminishes rapidly and becomes marginal for values above 0.0045. On the other hand, the real atomic column distribution has a positive median of 0.0052 and is skewed toward the right. Many values are seen exceed 0.010, which virtually never occurs for spurious atomic columns. The negative tail becomes negligible for values below -0.0060. These distributions reveal two simple guidelines: (1) sites with log-likelihood ratio values $\geq 0.0050$ can be treated as real structure with a high degree of certainty, and (2) sites with log-likelihood ratios $\leq -0.0060$ (e.g., the spurious column arrowed in **Figure 6c**) are almost certainly artificial. A site with a value in between is not as easily distinguishable but nonetheless still has a quantitative statistical measure of agreement given by its log-likelihood ratio. In practice, additional prior information (e.g., knowledge of the material) can also be leveraged to support an assessment of the predicted structure.

*3.5. Performance on Experimental Data and Visualizing the Network's Effective Filter*

The trained network was applied to the experimentally acquired *in situ* TEM image dataset. Several other state-of-the-art denoising techniques were also applied to the same real data in order to establish a baseline for evaluating the performance of the proposed network. **Figure 7** presents a summary of the results. A single 25 ms exposure *in situ* TEM image of a CeO$_2$-supported Pt nanoparticle in 5 mTorr N$_2$ gas is shown in **Figure 7a**. Beneath it, a zoom-in image is shown from the region marked by the red box at the Pt nanoparticle surface, to demonstrate the severity of the shot noise and the lack of clarity regarding the underlying image signal. Each baseline method was applied to the same noisy image, generating the denoised outputs shown from **Figure 7b** to **Figure 7g**. Details on all of the methods are given in **Section 2.4**. The denoised image produced by the proposed network architecture is shown in **Figure 7h**. Although a clean reference image is not available experimentally, a relatively high SNR image has been prepared by time-averaging the experimental data over 40 frames for 1.0 sec total, as shown in **Figure 7i**. Finally, **Figure 7h** displays the likelihood map for interpreting the structure that appears in the proposed network's output.

As seen in comparing the time-averaged image against the denoised estimates generated by the various methods, the proposed network architecture produces denoised images of superior quality. In particular, the proposed network is the only method that recovers a physically sensible atomic structure at the Pt surface, with the denoised zoom-in of **Figure 7h** strongly resembling the time-averaged zoom-in of **Figure 7i**. The DnCNN (**Figure 7f**) and small UNet (**Figure 7g**) denoising networks achieve the next-best overall performance. However, the images output by these architectures tend to exhibit unphysical structures characterized by, e.g., warped contrast around corner sites, not to mention that they also show unusual atomic column-like intensity in the vacuum and at the Pt surface, likely due to localized noise fluctuations. The remaining methods yield images of relatively similar inferior quality. A remarkable exception worth mentioning is the blind-spot network (**Figure 7b**). This self-supervised deep learning method, which was trained only on the raw experimental data and not on the simulations, outputs an image with arguably worse noise content in the image center around the Pt nanoparticle and Pt/CeO$_2$ interface; interestingly, in other regions (e.g., the vacuum and the CeO$_2$ bulk), the denoised estimate matches the time-averaged image contrast with exceptional similarity. We are presently investigating alternative blind-spot architectures for improved performance (Sheth et al., 2020). Another series of denoised images generated from another experimental frame is shown in **Supplemental Figure S10**.

The denoising mechanisms used by CNNs are often treated as a "black box", with little understanding offered to interpret how they work. Recent work shows that computing the gradient of the network's output with respect to its input at a specific pixel of interest can offer an interpretable visualization of the network's equivalent linear filter at that pixel (Mohan et al., 2020). In this section, we investigate the filtering strategies used by the network to denoise real data and show how they adapt to the presence of atomic-level defects at the catalyst surface.

Consider the denoised experimental frame shown in **Figure 8a**. Three pixels in the image have been marked by (small) red squares. One pixel is in the vacuum, one is in an atomic column at the Pt nanoparticle surface, and the last is in an atomic column in the CeO$_2$ bulk. The effective receptive field around each pixel is marked by a larger red box; these regions are plotted in **Figure 8(b1)**, **Figure 8(c1)**, and **Figure 8(d1)**, respectively, with the pixels of interest again marked by a small red square. It is noted that while the true receptive fields around each pixel are about 800 x 800 pixels in size, most of the information in the gradient is concentrated around the central 300 x 300 pixels, so for plotting purposes we choose to focus on this region. We wish to investigate the mechanism by which the network denoises these particular pixels. **Figures 8(b2)**, **(c2)**, and **(d2)** display the field of view around each pixel in the noisy experimental data. These windowed images are effectively what the network senses when denoising each pixel. In **Figures 8(b3)**, **(c3)**, and **(d3)** the Jacobian of the network at each pixel is plotted, which gives a local linear approximation of the function used by the network to map the noisy input to a denoised output. We call this visualization the network's effective filter, as it shows which regions of the input have the most impact on the denoised estimate.

Interestingly, the effective filter shows considerable variation at different locations in the image. For the pixel in the vacuum, **Figure 8(b3)** shows the gradient at this location is mostly uniform with a magnitude close to 0.0. The largely uniform gradient suggests the network senses a lack of structure in the vacuum and has incorporated this information into its denoising strategy. Compare this with the gradient plotted in **Figure 8(d3)** for the pixel on an atomic column in the $CeO_2$ bulk. Here, the gradient shows a clear periodicity, with a symmetric pattern that mirrors the local structure of the bulk material. The symmetry reveals that the network has learned to recognize an uninterrupted continuation of structure at this location. Note that the magnitude of the gradient in the region around the central pixel is comparable to that of the surrounding atomic column-like regions. The mostly equal weighting of local and non-local periodic information implies that the network considers the central atomic column to be similar to those surrounding it.

The network's denoising strategy adapts in response to non-periodic structural features at the Pt nanoparticle surface. As seen in **Figure 8(c3)**, at the surface the network gives substantially more weight to information that is in the immediate proximity of the pixel to be denoised. Strongly weighting the intensity within an atomic column-sized region may be what enables the network to recover the non-periodic atomic features at the catalyst surface. In unfavorable cases, the same strategy could lead to artifacts if the noisy input contains a randomly bright clustering of intensity that resembles an atomic column. As in the $CeO_2$ bulk, periodicity is seen in the gradient at the Pt surface, although now the separation distance between the atomic column-like regions has changed to match the periodicity of the projected Pt lattice. Notably, the spatial distribution of the filter is also now less symmetric, with the magnitude of the gradient diminishing to zero more rapidly in the regions that contain vacuum. Hence, the asymmetry reflects the termination of the nanoparticle structure and suggests that the network has learned to identify the presence of the catalyst surface.

## 4. Conclusion

A supervised deep convolutional neural network has been developed to denoise atomic-resolution TEM images of nanoparticles acquired during applications wherein the image signal is severely limited by Poisson shot noise. Multislice image simulations were leveraged to generate a large dataset images for training and testing the network. The proposed network outperforms existing methods, including other convolutional neural networks, by a PSNR of 12.0 dB, achieving a PSNR of about 43 dB on a test set of simulated images (the typical starting PSNR of the data explored in this work is only 3 dB). We show that the network is generally robust to ± 5 nm variations in defocus, although we suggest training the network using images at a defocus similar to the data that is to be denoised. The network's ability to correctly predict the atomic-scale structure of the nanoparticle surface was assessed by comparing the atomic columns originally present in clean simulations against those that appear in denoised images. We have also developed an approach based on the log-likelihood ratio test that provides a quantitative measure of the agreement between the noisy observation and the atomic-level structure present in the denoised image. The proposed assessment method requires only the network-denoised image and the noisy input and is therefore extensible to real experimental data, where no ground truth reference images exist. The network was applied to an experimentally acquired TEM image dataset of a $CeO_2$-supported Pt nanoparticle. We have conducted a gradient-based analysis to investigate the mechanisms used by the network to denoise experimental images. Here, this shows the network both (a) exploits information on the surrounding structure and (b) adapts its filtering approach when it encounters non-periodic terminations or atomic-level defects at the nanoparticle surface. The approaches described here may be applicable to a wide range of atomic resolution imaging applications that are characterized by ultra-low SNR, including the investigation of dynamic processes with time-resolved *in situ* microscopy or the study of beam sensitive systems.

## 5. Acknowledgements and Contributions to the Research

The authors gratefully acknowledge financial support from the National Science Foundation (NSF). NSF CBET Award 1604971 supported JLV and PAC, who acquired and processed the experimental data. NSF OAC Award 1940263 supported RM and PAC, who created the training dataset. NSF NRT HDR Award 1922658 partially supported SM who designed, implemented and analyzed the convolutional neural network and denoising methodology. NSF OAC Award 1940124 and NSF CCF Award 1934985 supported BT, who contributed to implement the methodology and conduct the experiments, and DM, who supervised the design of the methodology. NSF OAC Award 1940097 supported CFG who supervised the design, implementation and analysis of the methodology. DYS assisted with the implementation of the methodology; EPS supervised the design, implementation, and analysis of the methodology. The authors acknowledge HPC resources available through ASU, NYU, and Cornell, as well as the John M. Cowley Center for High Resolution Electron Microscopy at Arizona State University. The authors also gratefully acknowledge use of microscopy resources at the National Institute of Standards and Technology in Gaithersburg, MD, and in particular the helpful assistance and hospitality of Dr. Wei-Chang David Yang, Dr. Canhui Wang, and Dr. Renu Sharma.

## 6. Declaration of Competing Interests

The authors declare no competing interests.

Figures and figure captions:

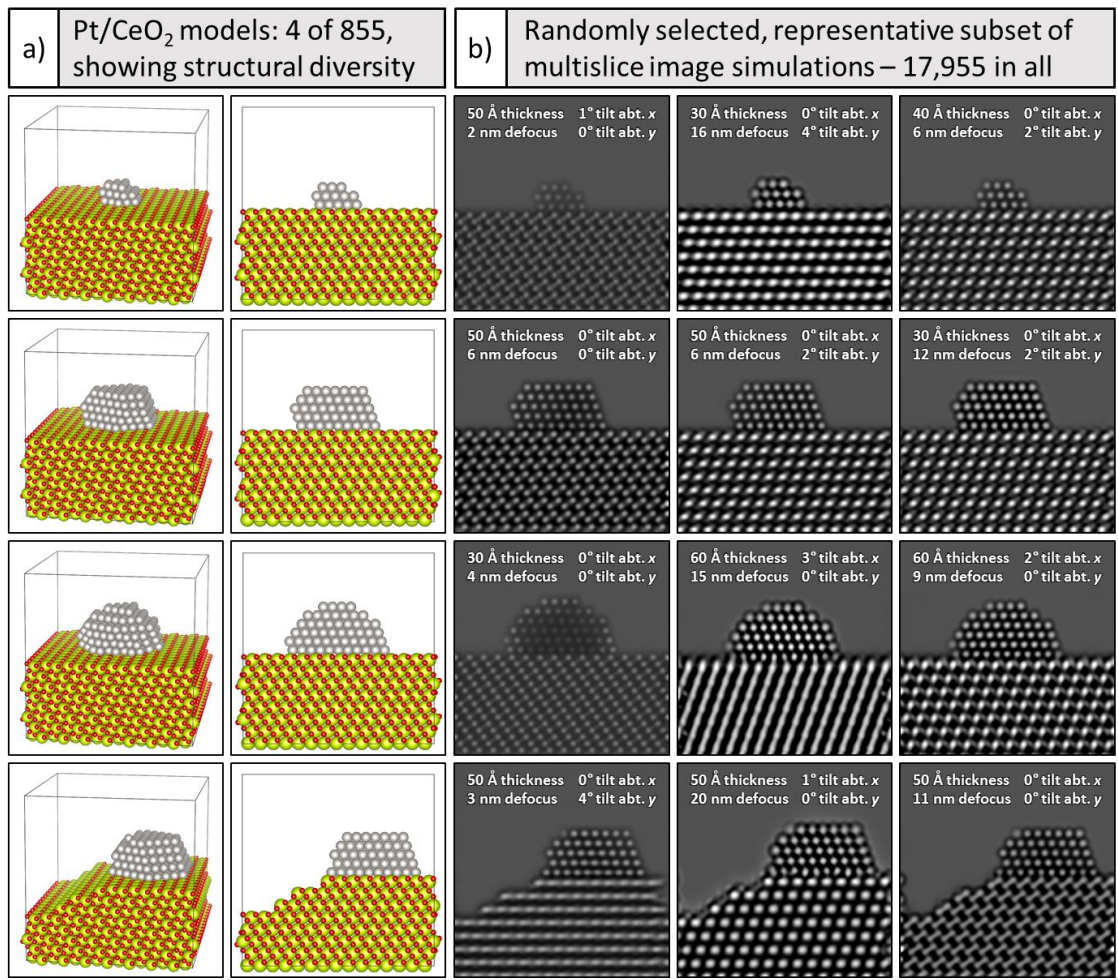

**Figure 1.** Generating a large training dataset through multislice image simulation. Under **(a)** four (of 855) models are shown in a tilted view to emphasize the 3D structure (far left) and in a projected view along the electron beam direction (second column). Pt atoms are shown in gray, O atoms in red, and Ce atoms in yellow-green. A simulated image of every structure was generated for defocus values spanning 0 – 20 nm, resulting in 17,955 total images. Beneath **(b)**, a representative subset of simulated images from each model is shown, with imaging conditions given in the figure inset (see text for more details).

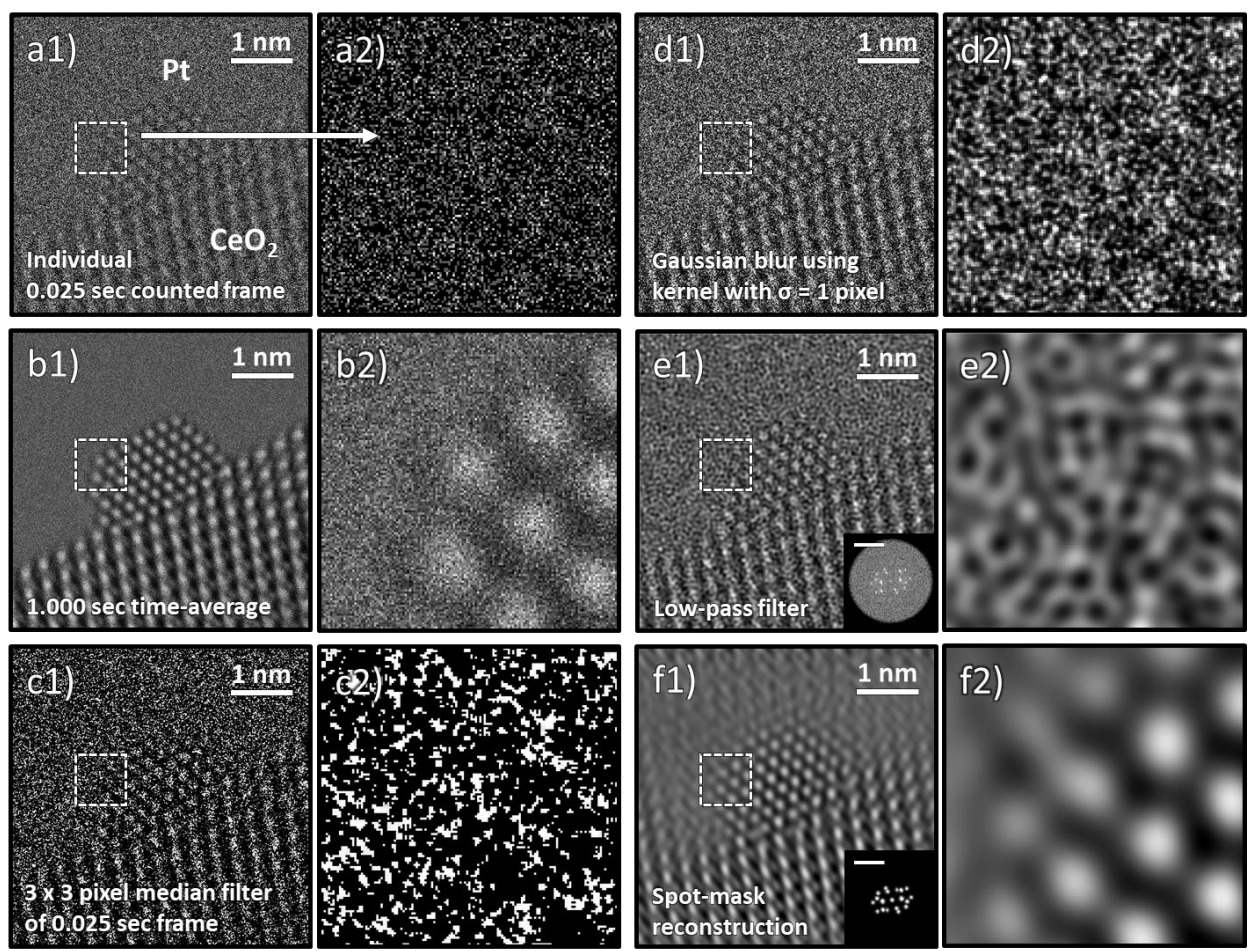

**Figure 2.** Comparison of typical processing techniques applied to an ultra-low SNR experimental TEM image of a $CeO_2$-supported Pt nanoparticle. In **(a1)** an individual 0.025 sec counted frame is shown along with **(a2)** a zoom-in image taken from the region designated by the dashed box. In **(b)** a 1.000 sec time-averaged image is shown; **(c)** displays the result of filtering the frame with a 3 × 3 pixel median filter; **(d)** displays the result of filtering the frame with a Gaussian blur with standard deviation equal to 1 pixel; **(e)** shows a Fourier reconstruction of the individual frame after applying a low-pass filter up to the 0.74 Å information limit, with the FT given in the inset along with a 1 Å$^{-1}$ scale bar; and **(f)** displays another Fourier reconstruction acquired through masking the Bragg beams in the diffractogram, as shown in the figure inset.

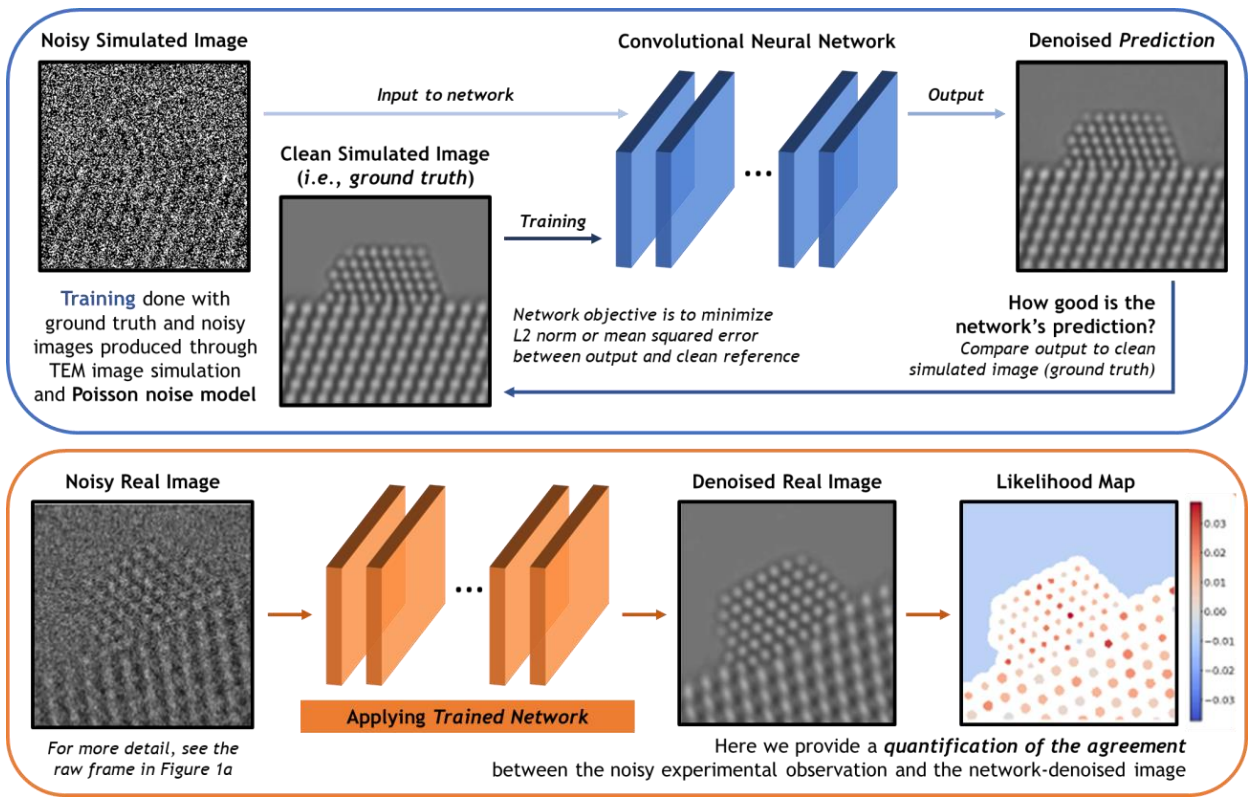

**Figure 3.** Overview of the deep convolutional neural network training, application, and evaluation process. **(Top)** The network is trained on a large dataset of noisy multislice TEM image simulations; the denoised prediction output by the network is compared to the original clean image simulation through a loss function based on the L2 norm (i.e., mean squared error). The parameters in the network are iteratively adjusted to minimize the magnitude of the loss function. **(Bottom)** The network trained on simulated images is then applied to real experimental data taken under similar imaging conditions. The performance of the network on real images lacking noise-free counterparts can be evaluated through a statistical likelihood analysis, which allows one to quantify the agreement between the denoised image and the noisy experimental observation.

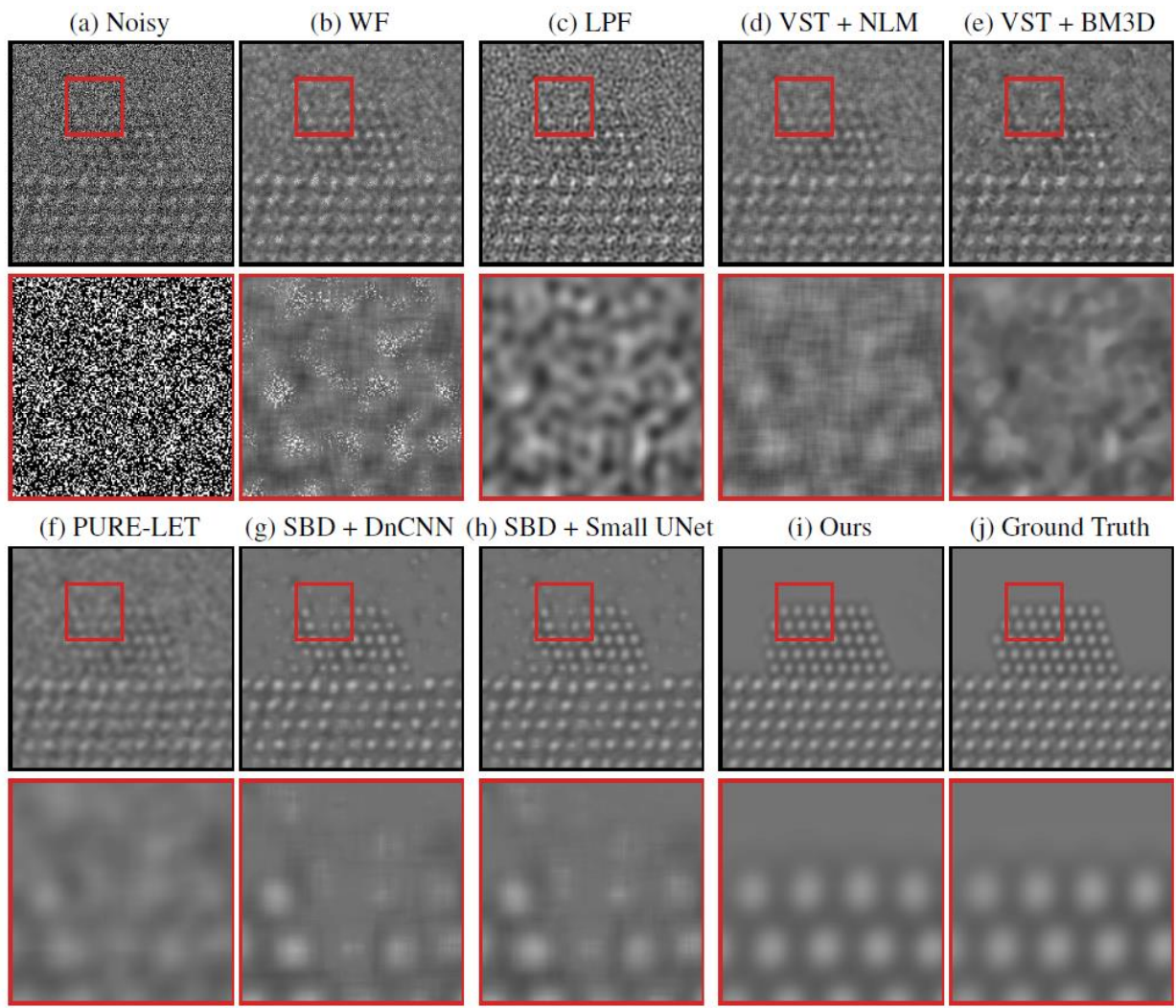

**Figure 4.** Comparing the proposed network's performance on multislice simulations against other baseline denoising methods, including other neural networks. See text for an explanation of the methods. In brief, part **(a)** displays a noisy simulated image, along with a zoom-in on the region indicated by the red box in the figure inset. The clean simulated image is shown as a ground truth reference in **(j)**. The proposed network produces denoised images of high quality, recovering precisely the structure of the nanoparticle, even at the surface, with comparatively few artifacts, as shown in **(i)**.

**Table 1**

Summary of denoising performance on simulated images in terms of mean peak signal-to-noise ratio (PSNR) and structural similarity (SSIM), along with the standard deviation, for each of the surveyed methods aggregated over all of the images in the validation dataset.

| Denoising Method | PSNR (dB) | SSIM (arb. units) |
|---|---|---|
| Raw | 3.56 ± 0.03 | 0.00 ± 0.00 |
| Adaptive Wiener Filter (WF) | 21.59 ± 0.07 | 0.44 ± 0.03 |
| Low-pass Filter (LPF) | 22.42 ± 1.08 | 0.63 ± 0.02 |
| VST + NLM | 26.55 ± 0.16 | 0.73 ± 0.01 |
| VST + BM3D | 22.57 ± 0.15 | 0.80 ± 0.01 |
| PURE-LET | 28.36 ± 0.88 | 0.93 ± 0.01 |
| SBD + DnCNN | 30.47 ± 0.64 | 0.93 ± 0.01 |
| SBD + Small UNet | 30.87 ± 0.56 | 0.93 ± 0.01 |
| Ours | 42.87 ± 1.45 | 0.99 ± 0.01 |

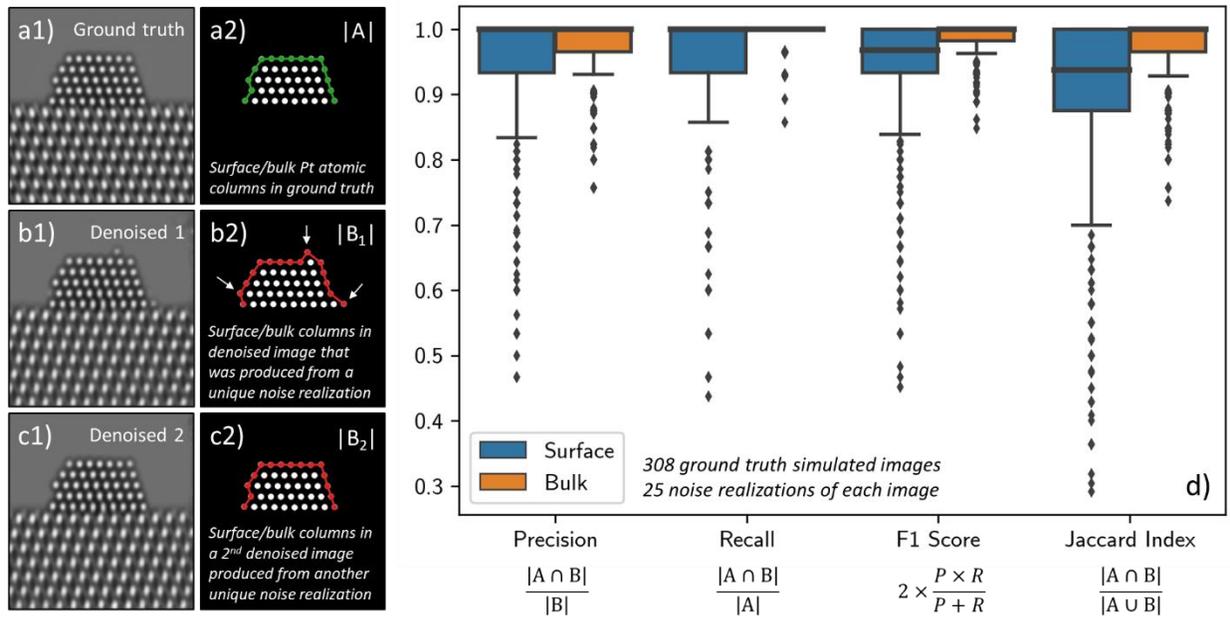

**Figure 5.** Part **(a1)** depicts a representative ground truth simulation from the Pt atomic structure evaluation image dataset ($n_{ground\ truth} = 308$). To the right, in **(a2)**, the set of Pt columns identified in the ground truth image (i.e., $|A|$) are shown, with those located at the surface highlighted by a green line. Parts **(b1)** and **(c1)** show two denoised images produced by the network from two unique noise realizations of the same original simulation. To the right, in **(b2)** and **(c2)**, the set of Pt columns identified in the respective denoised images (i.e., $|B|$) are shown, with those at the surface highlighted now by a red line. To quantify the network's performance in recovering the Pt atomic structure, we compute the precision, recall, F1 score, and Jaccard index of the two sets. Part **(d)** provides box plot distributions of each metric for both the surface (blue boxes) and the bulk (orange boxes) computed over 25 noise realizations of each ground truth simulation ($n_{denoised} = 7{,}700$). Outliers in the distributions are marked by small diamonds.

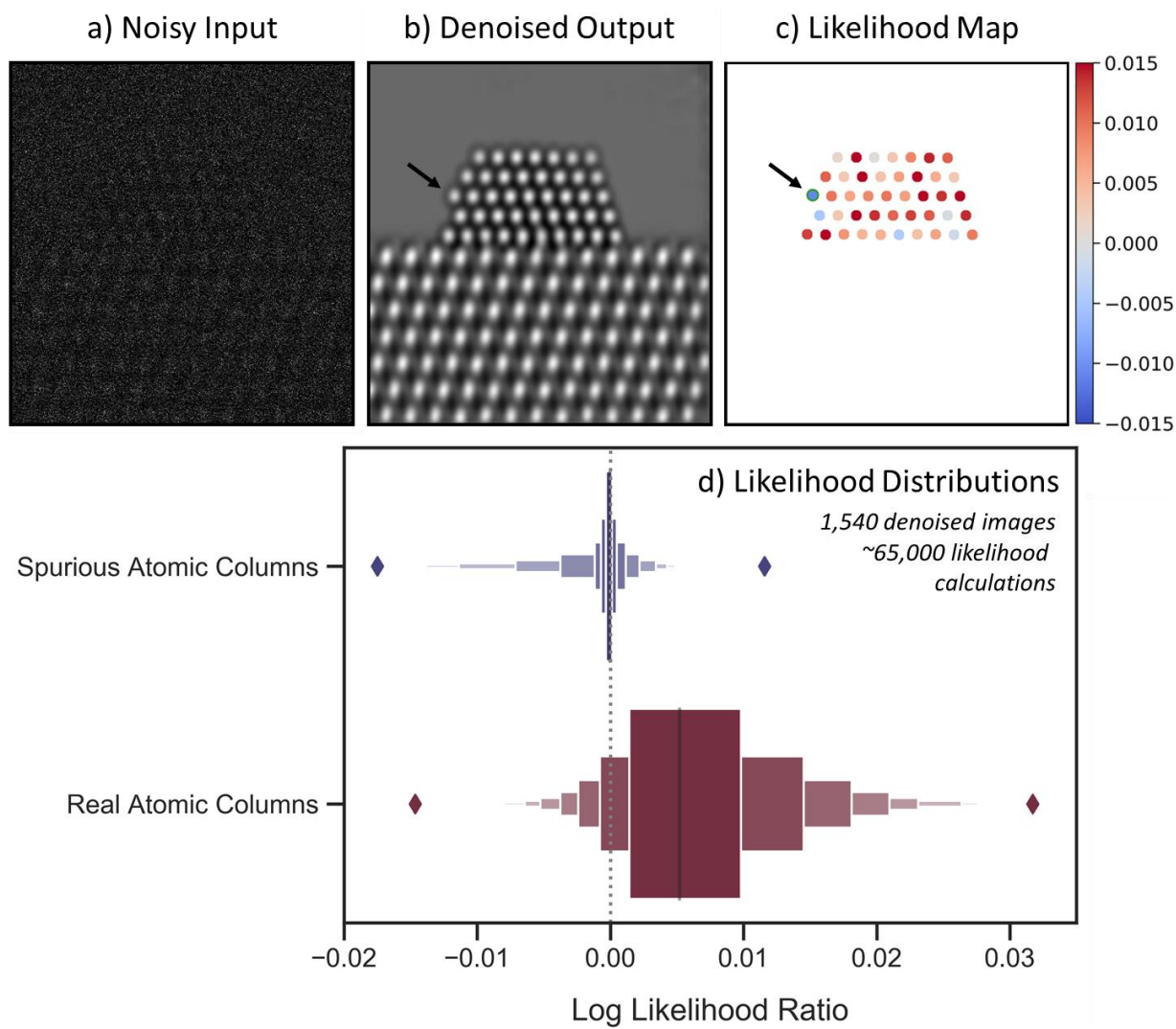

**Figure 6.** Likelihood analysis to quantify agreement between noisy data and network-denoised output. In **(a)** a representative noisy simulated image is shown along with **(b)** a denoised image output by the network. Part **(c)** depicts an atomic-level likelihood map, which visualizes the extent to which the atomic structure identified in the denoised image is consistent with the noisy observation. After denoising, a spurious atomic column appears at the arrowed site, which shows a large negative value in the likelihood map, indicating that the presence of an atomic column at this location is not likely. The likelihood analysis has been performed over 1,540 denoised images, yielding the distributions given by the letter-value plots for spurious (blue, top) and real (red, bottom) columns in part **(d)**. The diamonds mark the extrema of the two distributions.

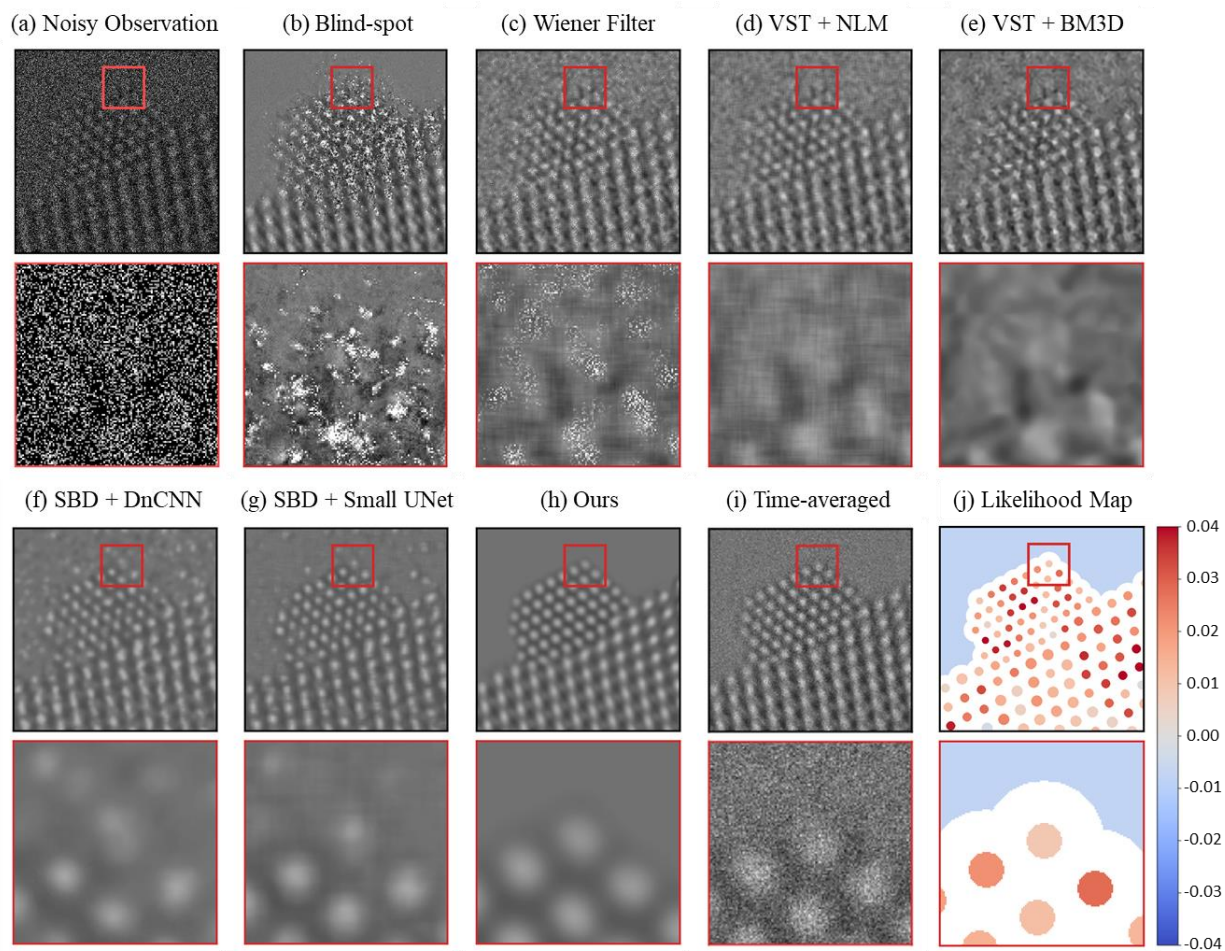

**Figure 7.** Evaluating the performance of the proposed network on experimental 25 ms exposure *in situ* TEM images, in comparison to current state-of-the-art methodologies. A raw 25 ms frame of a $CeO_2$-supported Pt nanoparticle in 5 mTorr $N_2$ gas is shown in **(a)** along with a zoom-in image from the region marked by the red box. Denoised estimates of the same raw frame from the baseline methods are presented in **(b)** through **(g)**, while **(h)** displays the denoised estimate from the proposed network. Part **(i)** presents a time-average over 40 raw frames, or 1.0 sec total, to serve as a relatively high SNR reference image. Finally, part **(j)** shows the likelihood map of the proposed network's output to quantify the agreement with the noisy observation.

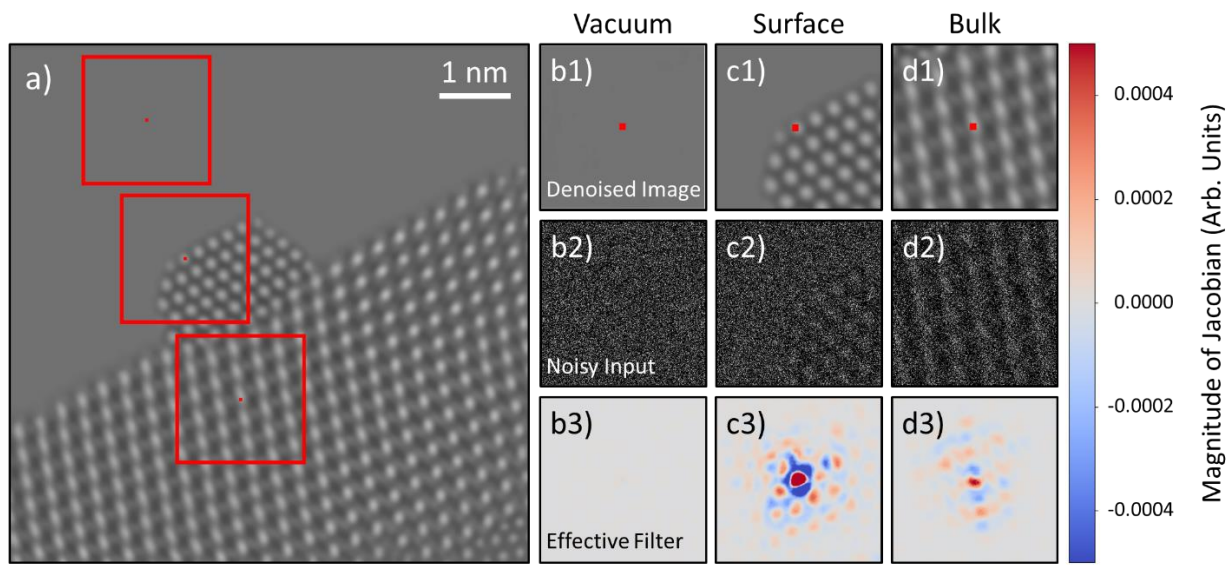

**Figure 8.** Investigating the mechanism by which the network denoises experimental data. A denoised experimental image is shown in **(a)**. Three regions of the image in the vacuum, catalyst surface, and bulk have been highlighted by red boxes and are depicted in **(b1)**, **(c1)**, and **(d1)**, respectively. The central pixel in each windowed region is marked with a red box. The noisy input within the network's receptive field around each pixel is displayed in **(b2)**, **(c2)**, and **(d2)**. In **(b3)**, **(c3)**, and **(d3)** the Jacobian of the network at each pixel is plotted, which provides an interpretable visualization of the regions of the noisy input that have the most impact on the denoised estimate.

*Supplemental Information*:

**Developing and Evaluating Deep Neural Network-based Denoising for Nanoparticle TEM Images with Ultra-low Signal-to-Noise**

Joshua L. Vincent[1], Ramon Manzorro[1], Sreyas Mohan[2], Binh Tang[3], Dev Y. Sheth[4], Eero P. Simoncelli[2,5,6], David S. Matteson[3], Carlos Fernandez-Granda[2,6]*, and Peter A. Crozier[1]*

[1] *School for Engineering of Matter, Transport, and Energy, Arizona State University, Tempe, AZ. 85287*

[2] *Center for Data Science, New York University, New York City, NY 10011*

[3] *Department of Statistics and Data Science, Cornell, Ithaca, NY 14850*

[4] *Indian Institute of Technology Madras, Chennai, Tamil Nadu 600036, India*

[5] *Center for Neural Science, New York University, New York City, NY 10011*

[6] *Courant Institute of Mathematical Sciences, New York University, New York City, NY 10011*

*Corresponding author emails: cfgranda@cims.nyu.edu and crozier@asu.edu

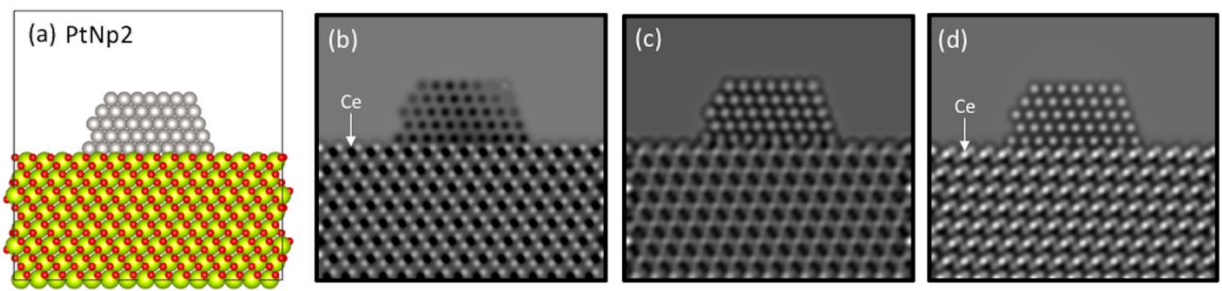

**Figure S1.** Categorical classification of "black", "intermediate", and "white" atomic-column contrast. The categorization was predominately centered around the focusing condition of the Pt atomic columns, with some influence as well by the focusing condition of the Ce atomic columns. In **(a)** an atomic-scale structural model of $CeO_2$-supported Pt is presented. Parts **(b)** through **(d)** show simulated images under different defocusing conditions, emphasizing variations in the Ce and Pt column contrast. In **(b)**, the image shows almost entirely black contrast for both Ce and Pt atomic columns. Images similar to this would be classified as "black" contrast. In **(c)**, the Pt columns reverse contrast and now appear white, while the Ce columns become challenging to discriminate. Images similar to this would be classified as "intermediate" contrast. Finally, in **(d)** all of the atomic columns including the O appear with white contrast. Images similar to this one would be classified as "white" contrast.

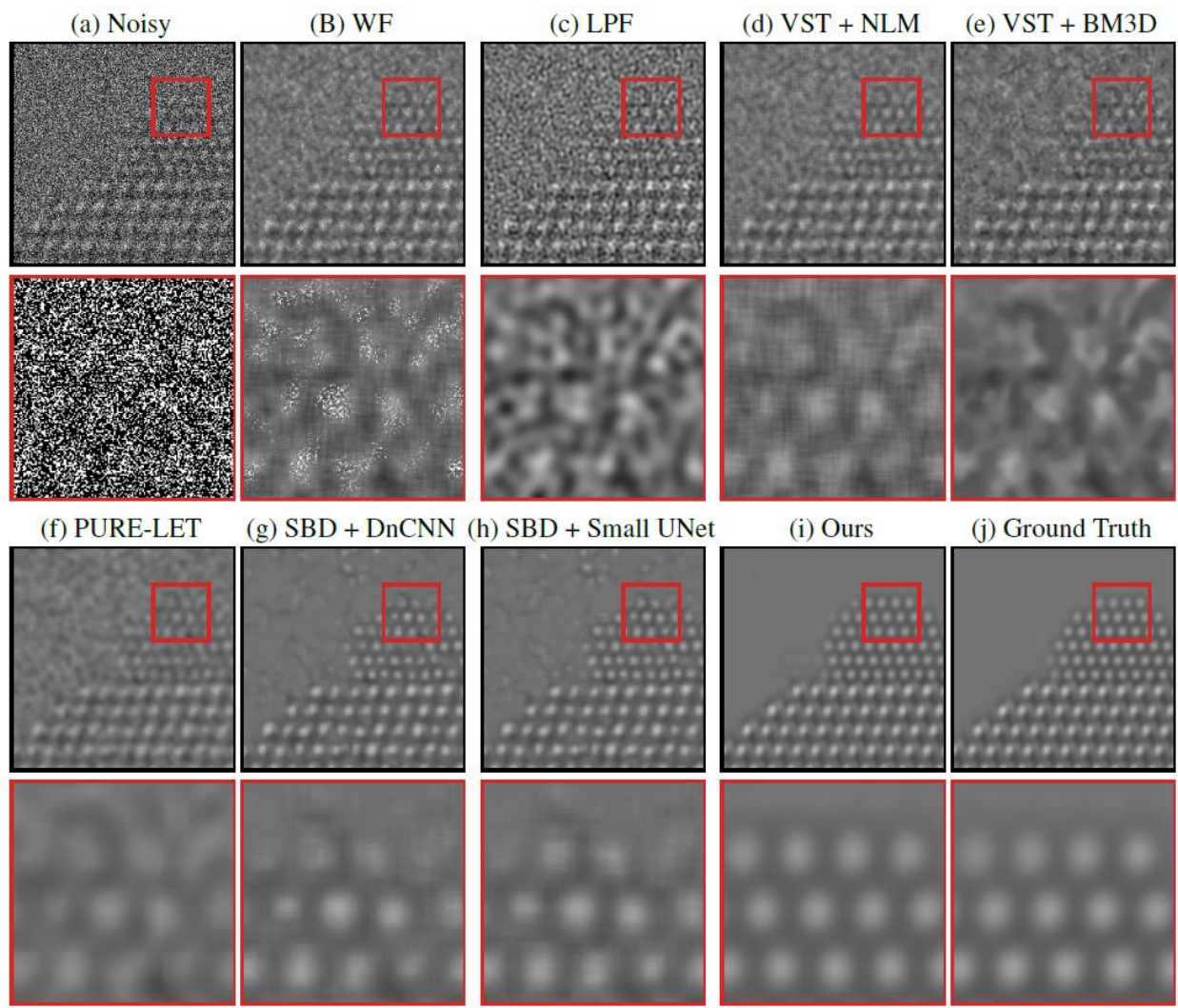

**Figure S2.** Comparing the proposed network's performance on a randomly selected simulated image from the validation dataset against other baseline denoising methods, including other neural networks. See main text for an explanation of the methods. In brief, part **(a)** displays a noisy simulated image, along with a zoom-in on the region indicated by the red box in the figure inset. The clean simulated image is shown as a ground truth reference in **(j)**. The proposed network produces denoised images of high quality, recovering precisely the structure of the nanoparticle, even at the surface, as shown in **(i)**.

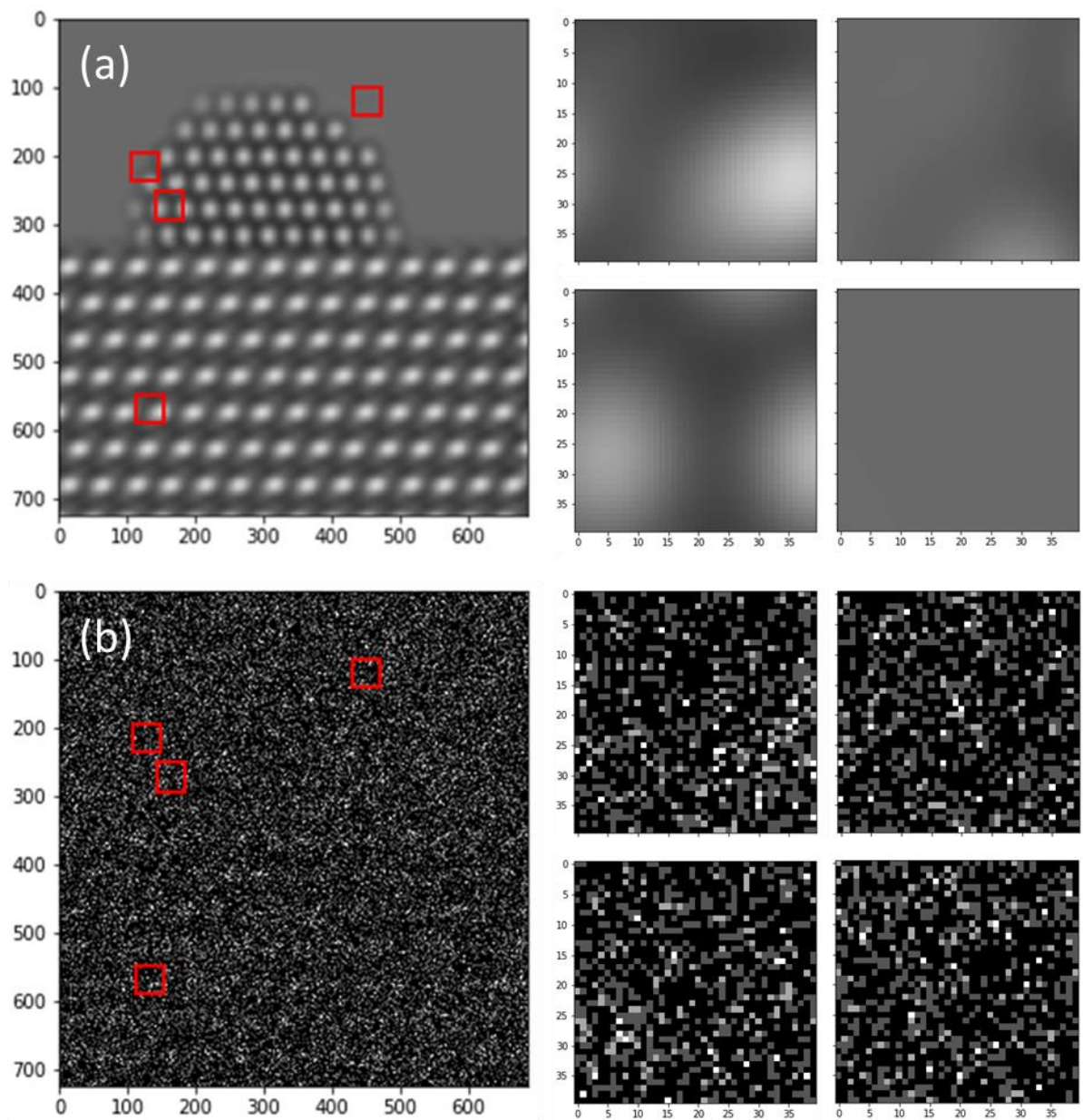

**Figure S3.** With a receptive field of 41 × 41 pixels, it is challenging to see structure around the atomic columns in the clean image, which is shown in (**a**) with randomly selected 41 × 41 pixel regions shown at right. After severely degraded Poisson shot noise has been added to the image, as shown below in (**b**), differentiating the regions which contain structure from those which are taken from the vacuum becomes considerably difficult.

Page 53

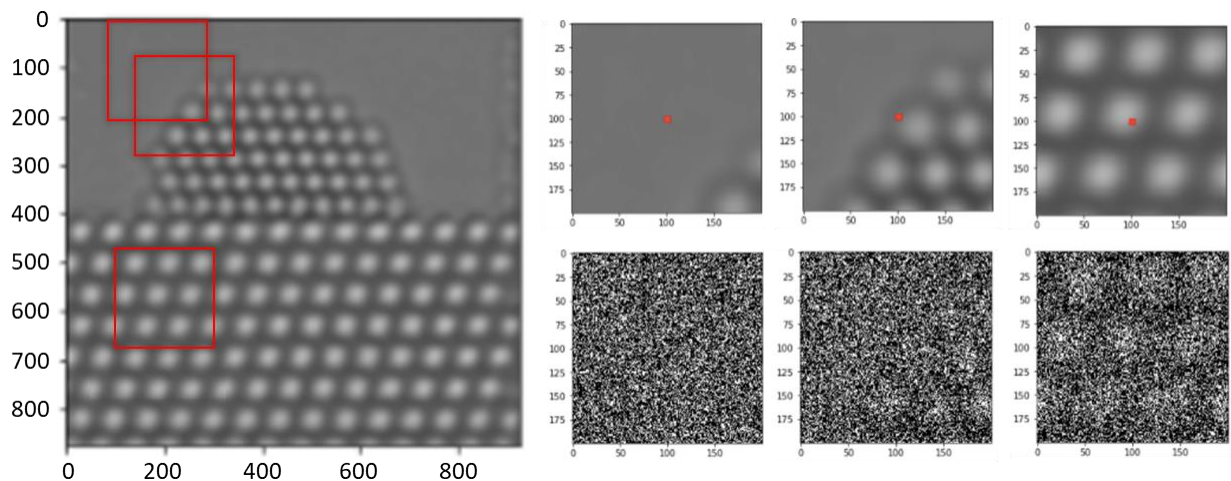

**Figure S4.** Increasing the network's receptive field (e.g., here regions 1.22 nm × 1.22 nm are shown) allows the network to sense nearby atoms, while remaining sensitive to the presence of a surface or defected site. Various regions of interest are highlighted by the red boxes in the image on the right. The local structure surrounding the pixel to be denoised (small red box in windowed regions shown on top right) can clearly be seen and remains discernible after the addition of severe shot noise (bottom right).

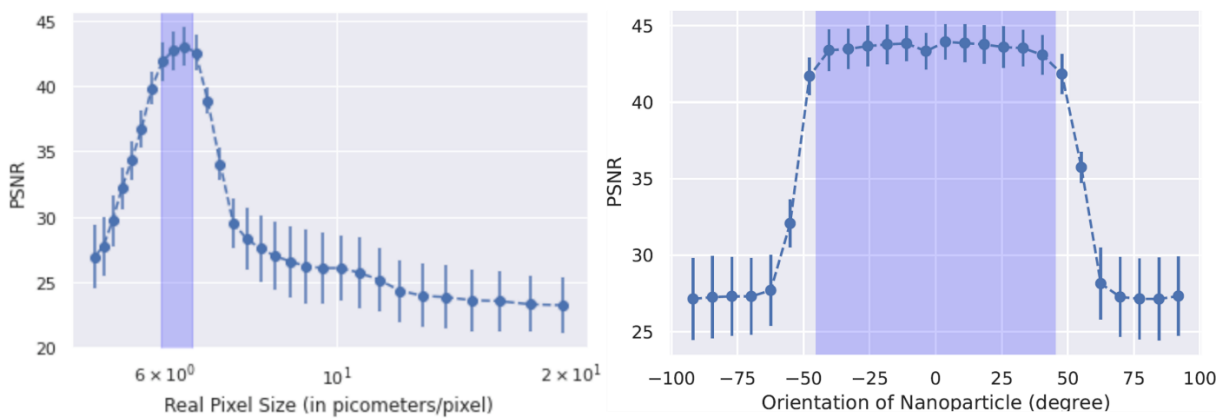

**Figure S5.** Impact of training data geometry on network denoising performance. At left, the effect of image scaling (measured in terms of real-space pixel size) is investigated; at right, the influence of image orientation (measured in degrees relative to the original simulation). In both cases the network was trained on data augmented with resized and rescaled images within the regions that are shaded purple. When the network is evaluated on images outside of these regions, the performance, measured in terms of PSNR, worsens significantly. Mean values are plotted for each size/orientation, with the standard deviation of the values given as the data error bars.

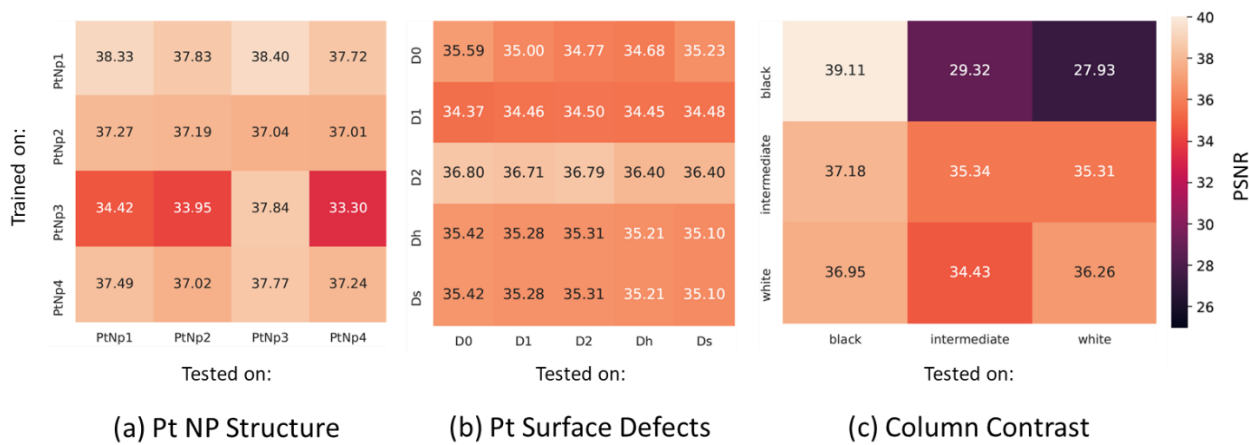

**Figure S6.** Investigating the network's generalizability to unseen **(a)** supported nanoparticle structures (see Figure S10), **(b)** atomic-level Pt surface defects (see Figure S11), and **(c)** atomic column contrast (i.e., white or black-column focusing) conditions (see Figure S1). A description of the different subsets of data that were formed for each category, as well as an explanation of the terminology, is given in the methodological section of the main text. The tables report the mean PSNR denoising performance when it is trained (rows) and evaluated (columns) on various combinations of the data subsets. For example, when the network is trained only on images with the PtNp1 structure (Table (a), row 1), the network achieves a PSNR denoising performance of 38.33 dB when it is evaluated on images of the PtNp2 structure, and a PSNR denoising performance of 37.72 dB when evaluated on images of the PtNp4 structure.

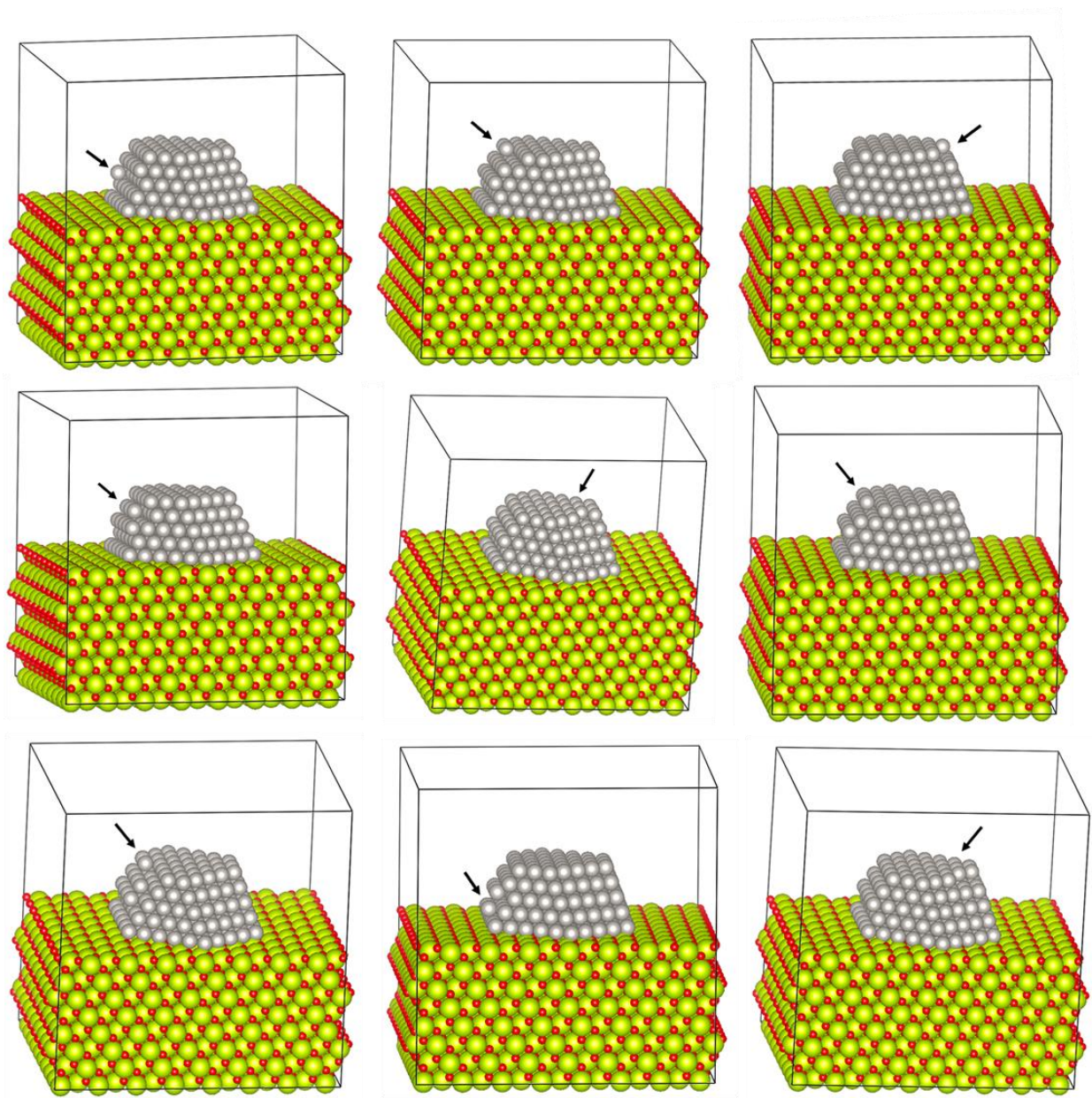

**Figure S7.** Representative set of nine Pt/CeO$_2$ atomic structural models used in the generation of the surface evaluation dataset. Many different types of atomic-level surface defects have been introduced into the Pt models, including, e.g., the removal of an atom from a column, the removal of two atoms, the removal of all but one atom, the addition of an adatom at a new site, etc., to emulate dynamic atomic-level reconfigurations that could potentially be observed experimentally. Altered sites are indicated with black arrows.

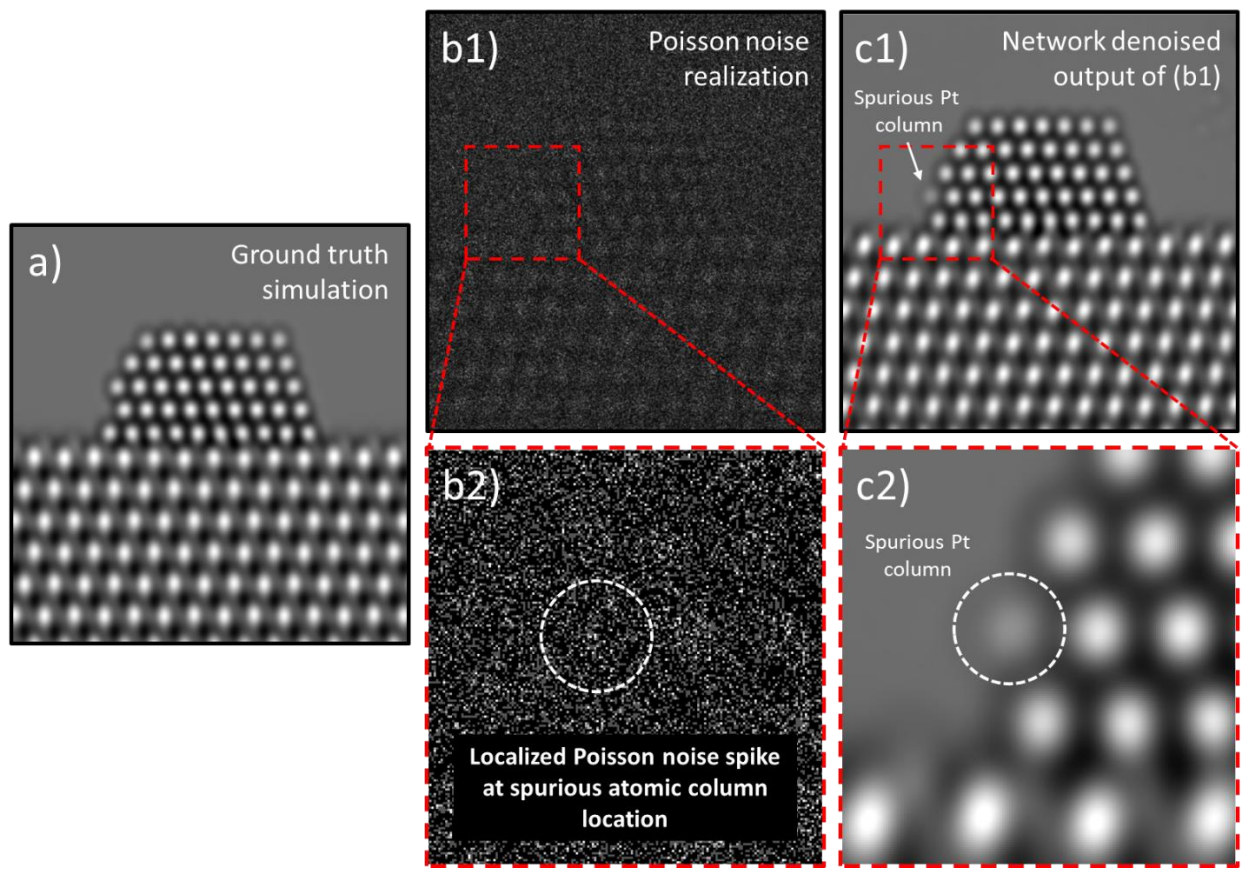

**Figure S8.** Examination of Poisson noise distribution around locations where spurious atomic columns appear in denoised simulated images. Part **(a)** shows the original ground truth simulation in this case, **(b1)** shows the Poisson noise realization and **(c1)** shows the network denoised output. Notice the appearance of a spurious atomic column which is marked by the white arrow in **(c1)**. Subfigures **(b2)** and **(c2)**, respectively, show an enhanced view around the spurious atomic column from the windowed region marked by the dashed red box in the noisy and denoised images. In **(b2)** and **(c2)** a dashed white circle is used to mark the location of the spurious atomic column. Examining the distribution of intensity in the Poisson shot noise realization reveals the presence of a noise spike near the center of the spurious atomic column location (i.e., **(b2)**). This analysis suggests that the random clustering of intensity in a manner that appears to resemble a surface atomic column can lead the network to produce denoised estimates with spurious surface atomic columns.

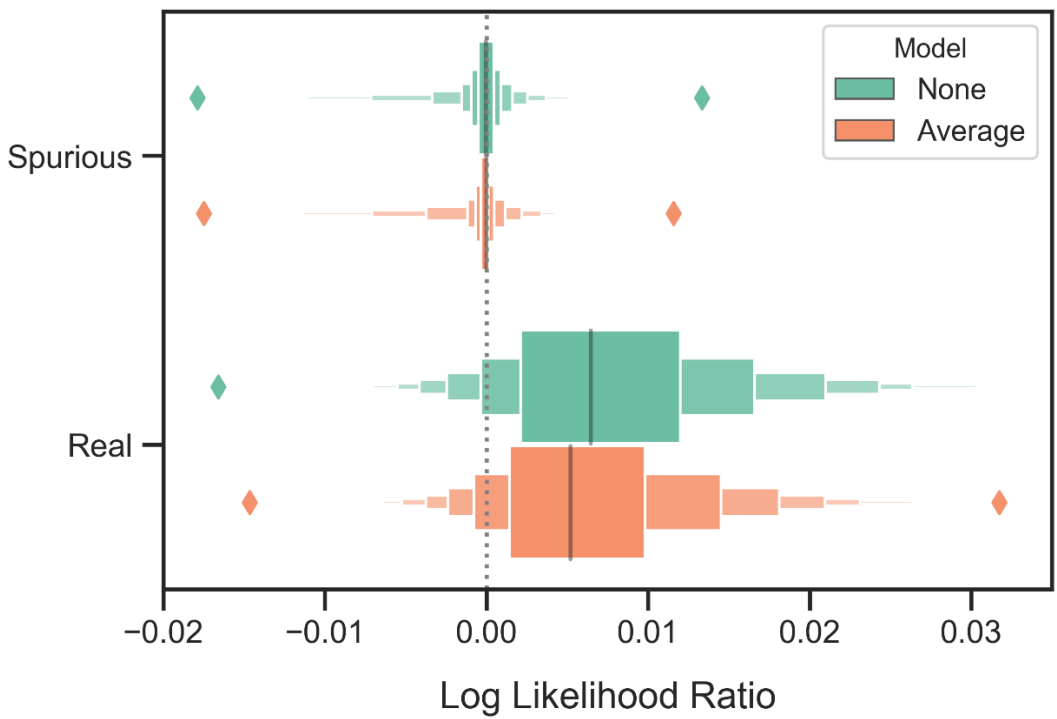

**Figure S9.** Letter value or so-called boxen plots of the log-likelihood ratio distributions for spurious (top) and real (bottom) atomic columns calculated in two different ways. The distribution labeled as having a model of "Average" contains log-likelihood ratios calculated using a Poisson probability mass function (pmf) governed by a rate parameter that was obtained by *averaging* the intensity within the column, as explained in the main text. The area over which the intensity was averaged is defined by a circle that is centered on the atomic column and approximately 1.5 Å in diameter. The distribution labeled as having a model of "None" contains ratios calculated using a Poisson pmf where the rate parameter of each pixel varies and is taken to be the intensity value of the denoised pixel. Observe that the distributions differ by little.

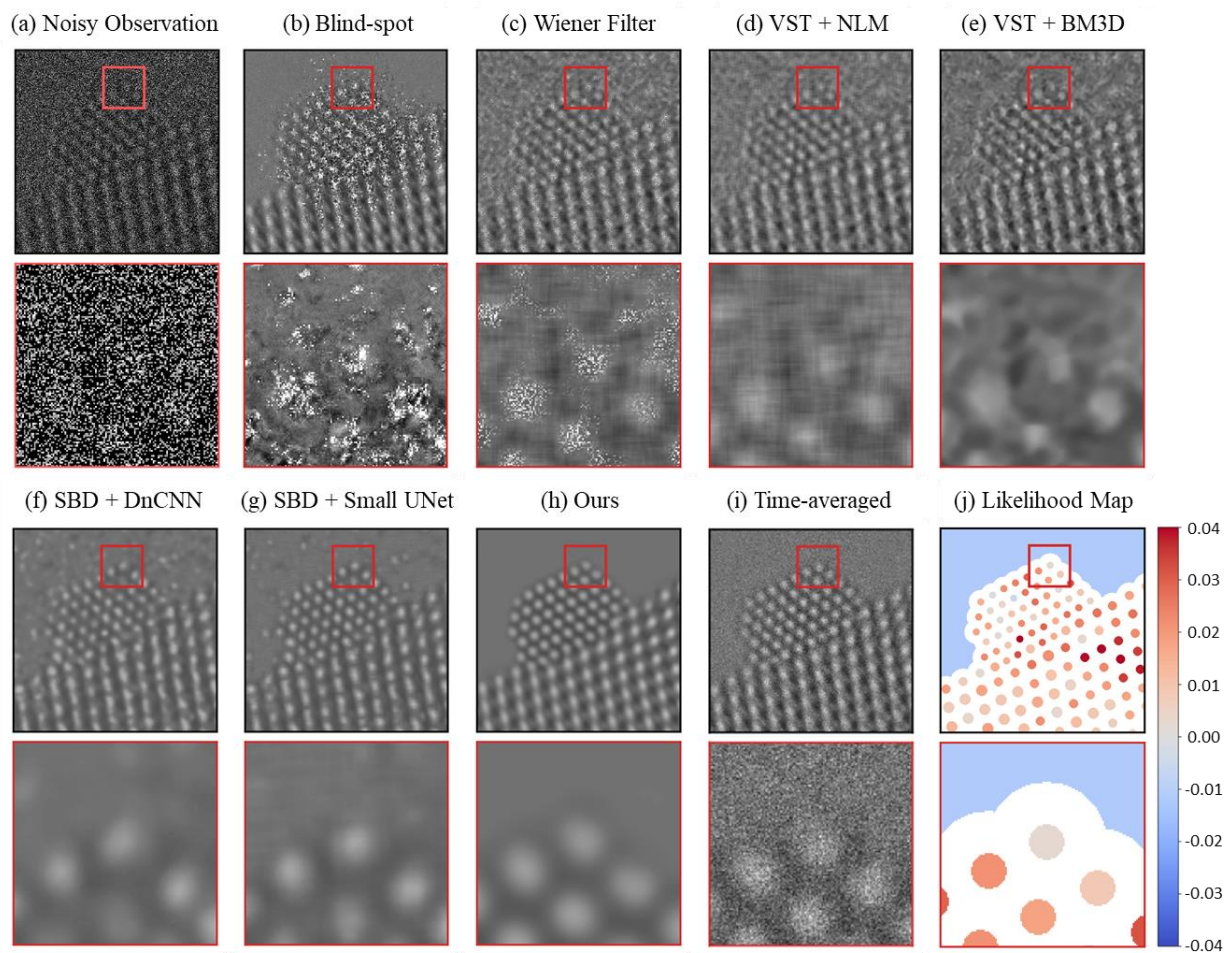

**Figure S10.** Evaluating the performance of the trained network on experimental 25 ms exposure *in situ* TEM images, in comparison to current state-of-the-art methodologies. A raw 25 ms frame is shown in **(a)** along with a zoom-in image from the region marked by the red box. Denoised estimates of the same raw frame from the baseline methods are presented in **(b)** through **(g)**, while **(h)** displays the denoised estimate from the proposed network. Part **(i)** presents a time-average over 40 raw frames, or 1.0 sec total, to serve as a relatively high SNR reference image. Finally, part **(j)** shows the likelihood map of the proposed network's output to quantify the agreement with the noisy observation.

**Appendix A: Description of Structural Variation Included in Atomic Models**

Four base supported Pt nanoparticle structures were incorporated in the model dataset to cover variations in the overall supported particle size and shape. The nanoparticle structures have been labeled "PtNp1" through "PtNp4", as shown below in **Figure S11**.

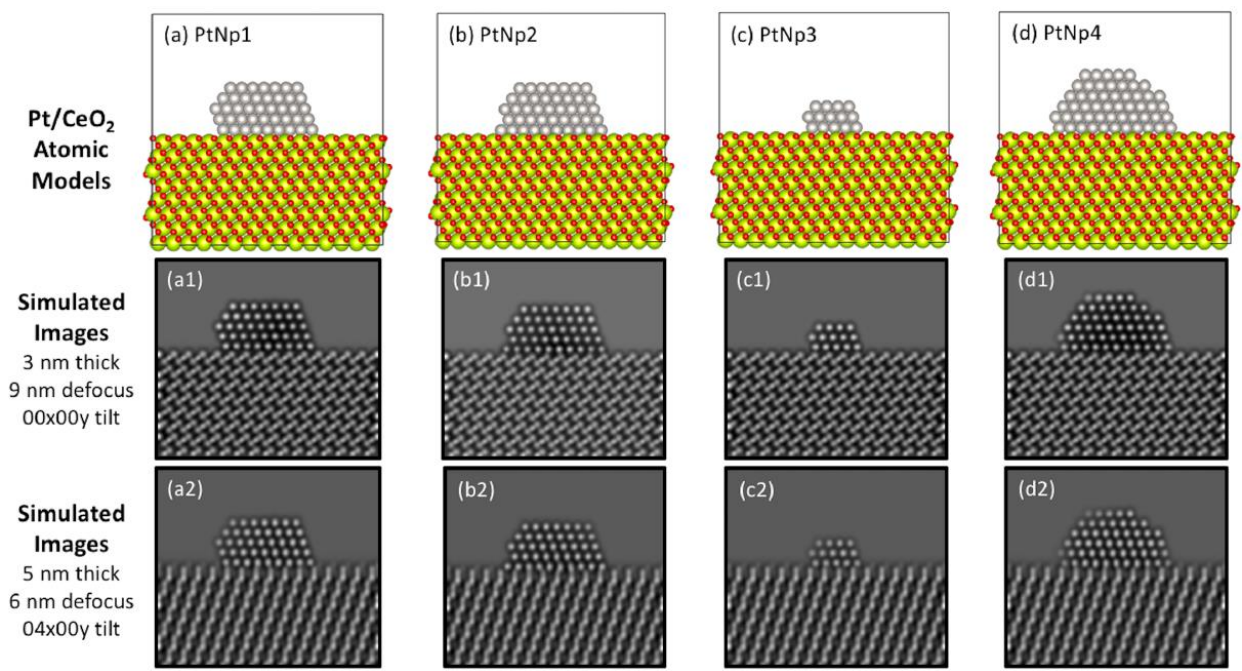

**Figure S11.** Variations in the size/shape of the supported Pt nanoparticle. At top, in **(a)** to **(d)**, atomic models of Pt nanoparticles PtNp1 through PtNp4, each with different size and shape, are supported on a $CeO_2$ slab. PtNp1 and PtNp2 correspond to supported Pt nanoparticles 2 nm in size where the difference is the appearance of an atomic column located at the interface between the Pt and the $CeO_2$ support; PtNp3 corresponds to a Pt nanoparticle 1 nm in size; and PtNp4 corresponds to a Pt nanoparticle 3 nm in size. In middle, from **(a1)** to **(d1)**, simulated images of the modeled structures are given for a $CeO_2$ support thickness of 3 nm, 9 nm of defocus, and no tilt; at the bottom, in **(a2)** to **(d2)** simulations for the same models are given now for a 5 nm support thickness, 6 nm of defocus, and 4° of tilt about the *x* axis.

Furthermore, the surface character of the Pt nanoparticles was varied by introducing atomic-level defects into the structure at different surface sites. A few examples are depicted below in **Figure S12**. Overall, the defects can be categorized into five classes, here labeled as "D0", "D1", "D2", "Dh", and "Ds" in accordance with the models presented below in **Figure S12**. In regard to the terminology, D0 corresponds to the initial structure without any introduced defects, D1 and D2 correspond to a structure in which 1 or 2 atomic columns have been removed, respectively, Dh corresponds to a structure in which a column has been reduced to half its original occupancy, and finally Ds corresponds to a structure in which a column has been reduced to a single atom. Note that the surface sites altered in the structure correspond to high-energy sites (e.g., corners and edges) which are more likely to dynamically rearrange or show variation than, say, a low-energy terrace site located in the middle of the surface.

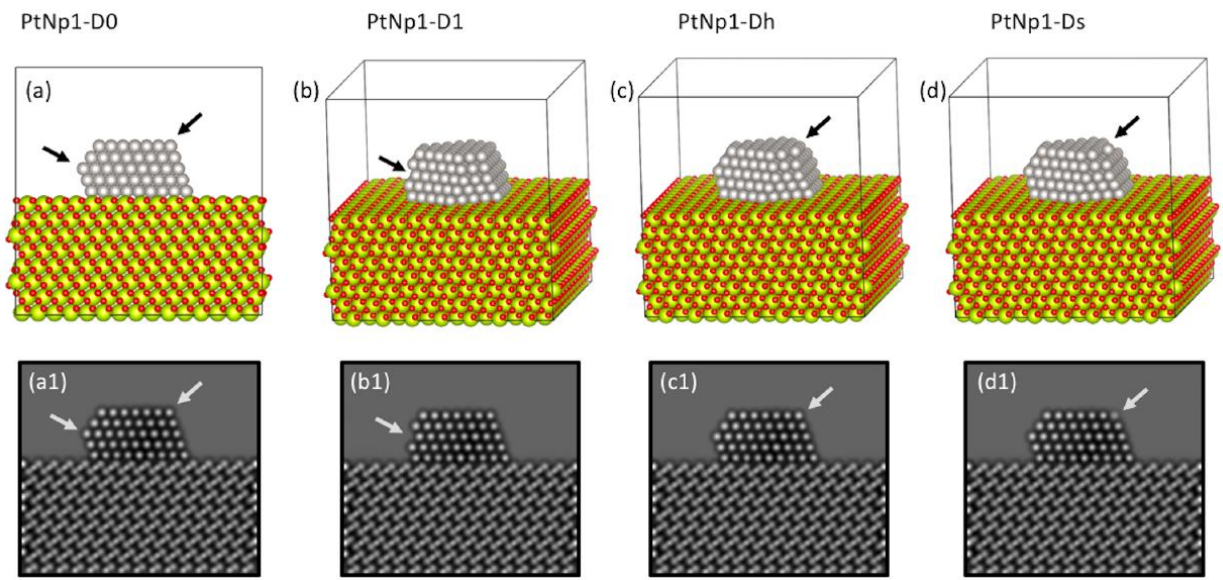

**Figure S12.** Variations in the atomic-level defects present on the Pt surface. In **(a)** an atomic model of CeO$_2$-supported Pt nanoparticle PtNp1, without any introduced defects (D0) is shown. The surface of this nanoparticle has been modified in a number of ways, including **(b)** by removing a full atomic column (i.e., defect D1), **(c)** by removing half of the column occupancy (defect Dh), and **(d)** by removing all but a single Pt atom (defect Ds). Black arrows point to the sites where the defects have been introduced. Note that models **(b)**, **(c)**, and **(d)** have been slightly tilted to assist in visualizing the surface defect modifications. At bottom in **(a1)** to **(d1)**, simulated images of the atomic models are shown for conditions with 3 nm support thickness, 9 nm of defocus, and no tilt.

Finally, the support thickness was varied from 3 nm to 6 nm along 1 nm increments. Images showing the type of contrast variations that may occur when the support thickness is changed, and how these compare to that which arises from changes in defocus, are given in **Figure S13** below.

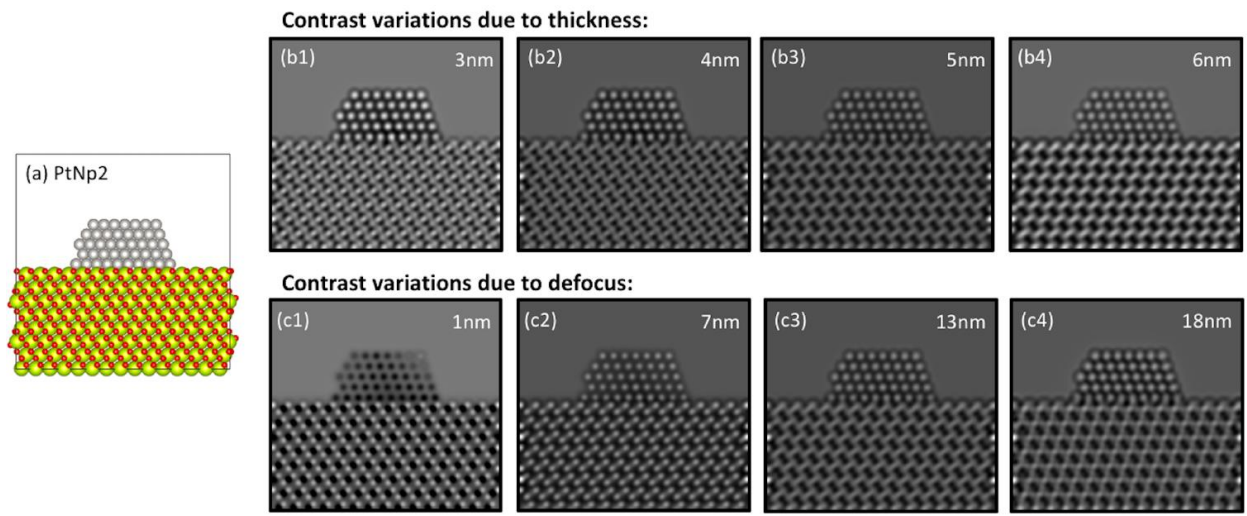

**Figure S13.** Image contrast variations due to $CeO_2$ support thickness (top) and electron optical defocus (bottom). The model shown in **(a)** was used for each of the multislice simulations to isolate effects from thickness and defocus. Images **(b1)** through **(b4)** demonstrate the effect of $CeO_2$ support thickness on the contrast in the image, with the thickness increased from 3 nm to 6 nm in 1 nm increments and the defocus held constant at 13 nm. Images **(c1)** through **(c4)** illustrate the effect of defocus on image contrast, with the defocus increased from 1 nm to 7 nm, then 13 nm, then 18 nm, respectively, and the support thickness held constant at 5 nm.

Aside from this, the overall orientation of the structural model with respect to the incident electron beam was tilted from 0° to 4° about the *x* and *y* axes independently in increments of 1°. Thus, variations from 0° in *x* and 0° in *y*, to 4° in *x* and 0° in *y*, or 0° in *x* and 4° in *y* were considered. Accounting for the diversity in structures, in addition to the variations in crystal orientation and $CeO_2$ support thickness, a total of 855 atomic structural models were constructed. These structures were each used to calculate multislice simulations with defocus values ranging from 0 to 20 nm, which results in the calculation of 17,955 total images.

**Figure S14** presents a schematic summary of the structural and imaging parameters varied during the modeling and image simulation process.

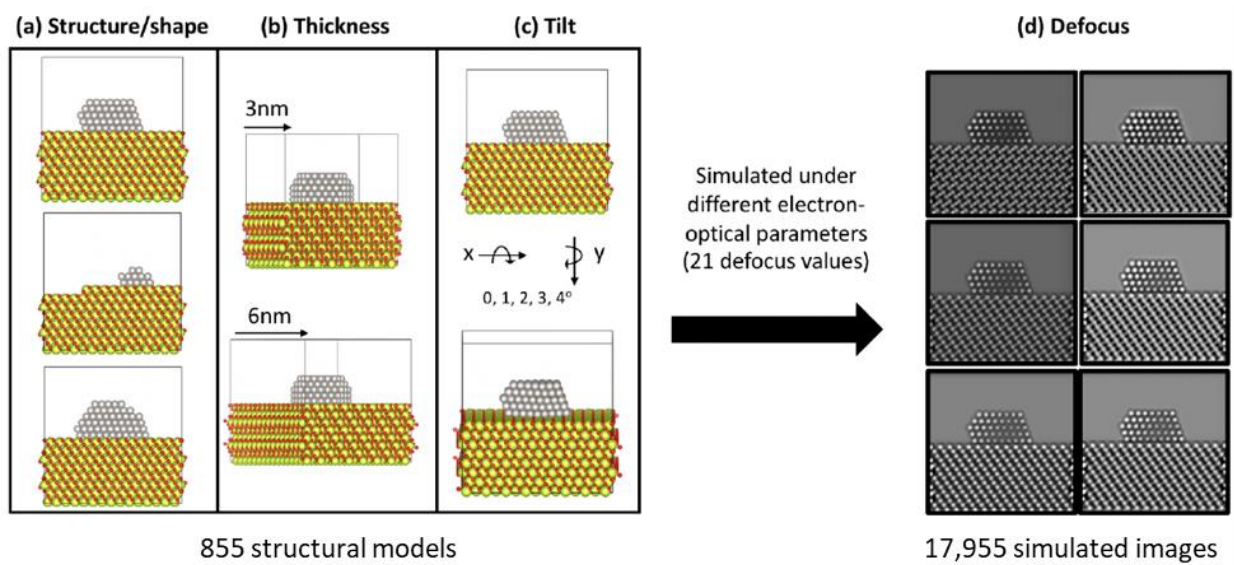

**Figure S14.** Summary of systematically varied structural and imaging parameters considered during the modeling and image simulation process. At left is a subset of Pt/CeO$_2$ atomic structural models presenting variations on the **(a)** structure and shape of the nanoparticle and the support, **(b)** the thickness of the CeO$_2$ support, and **(c)** the tilt of the specimen with respect to the incident beam. The models were used to produce simulations under 21 defocus values each, as shown in **(d)**.

**Appendix B: Analysis of Experimental Noise Distribution**

Given the physical origin of the noise in the experimental image acquisition process, we expect the noise to be dominated by shot noise, which can be modeled with a Poisson distribution. Here, the images were acquired on a direct electron detector operating in electron counting mode. In such conditions, the electron dose rate per pixel is sufficiently low enough that individual electron arrivals can be detected and registered. It is well known that the statistical fluctuations of such counting processes for discrete events are governed by shot noise. Additionally, we expect that other sources of noise, including fixed pattern noise, dark noise, and thermal noise are minimal after applying a gain correction and a dark reference to the raw image, and by cooling the detector to -20 °C, respectively. Readout noise is considered to be negligible, since the pixels on the CMOS-based detector are read out individually. Thus, we expect that the noise in the counted TEM micrographs can be modeled as Poisson. Furthermore, we have performed an analysis to verify that the noise in the experimental movie follows Poisson statistics, as shown below in **Figure S15**.

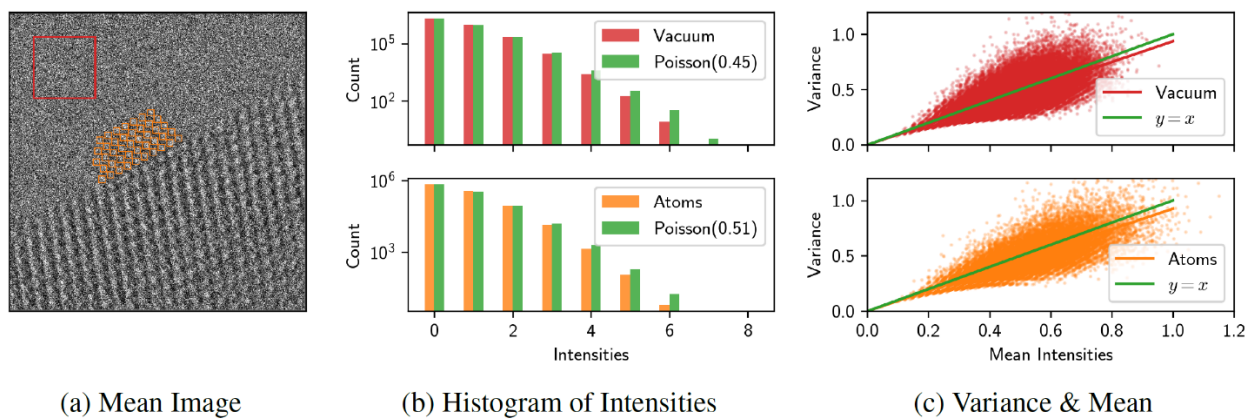

(a) Mean Image  (b) Histogram of Intensities  (c) Variance & Mean

**Figure S15.** In **(a)** a 1.000 second time-averaged image comprised of 40 frames is displayed. Part **(b)** displays histograms from the red and orange regions in the image representing vacuum and the Pt atomic columns, respectively. Simulated histograms taken from Poisson distributions with the indicated mean are plotted for comparison, showing good agreement in both cases. Finally, in **(c)** a plot of the mean and standard deviation of the pixel intensities over the 40 frames in the movie shows the data approximately follows a line with a slope of 1, as expected for Poisson distributions. The spread in the data is due to the limited number of samples (i.e., 40).